\documentclass[aps,prb,twocolumn,floats,epsfig,pdflatex]{revtex4}

\usepackage{epsfig}
\usepackage{amsmath}
\usepackage{amsfonts}
\usepackage{graphicx}
\usepackage{amssymb}
\usepackage{amsbsy}
\usepackage{color}

\newcommand{\beq}{\begin{equation}}
\newcommand{\eeq}{\end{equation}}
\newcommand{\bea}{\begin{eqnarray}}
\newcommand{\eea}{\end{eqnarray}}

\begin{document}

\title{A Floquet perturbation theory for periodically driven weakly-interacting fermions }
\author{Roopayan Ghosh, Bhaskar Mukherjee, and K. Sengupta}
\affiliation{School of Physical Sciences, Indian Association for the
Cultivation of Science, Kolkata 700032, India.}
\date{\today}

\begin{abstract}

We compute the Floquet Hamiltonian $H_F$ for weakly interacting
fermions subjected to a continuous periodic drive using a Floquet
perturbation theory (FPT) with the interaction amplitude being the
perturbation parameter. This allows us to address the dynamics of
the system at intermediate drive frequencies $\hbar \omega_D \ge V_0
\ll {\mathcal J}_0$, where ${\mathcal J}_0$ is the amplitude of the
kinetic term, $\omega_D$ is the drive frequency, and $V_0$ is the
typical interaction strength between the fermions. We compute, for
random initial states, the fidelity $F$ between wavefunctions after
a drive cycle obtained using $H_F$ and that obtained using exact
diagonalization (ED). We find that FPT yields a substantially larger
value of $F$ compared to its Magnus counterpart for $V_0\le \hbar
\omega_D$ and $V_0\ll {\mathcal J}_0$. We use the $H_F$ obtained to
study the nature of the steady state of an weakly interacting
fermion chain; we find a wide range of $\omega_D$ which leads to
subthermal or superthermal steady states for finite chains. The
driven fermionic chain displays perfect dynamical localization for
$V_0=0$; we address the fate of this dynamical localization in the
steady state of a finite interacting chain and show that there is a
crossover between localized and delocalized steady states. We
discuss the implication of our results for thermodynamically large
chains and chart out experiments which can test our theory.

\end{abstract}


\maketitle

\section{Introduction}
\label{intro}

The study of non-equilibrium dynamics of correlated quantum systems
has seen tremendous progress in recent years \cite{rev1,rev2}. Out
of several possible protocols of driving such systems, periodic ones
lead to several interesting phenomena that have no analogues for
their aperiodic counterparts \cite{rev3}. Some of these phenomena
include realization of novel steady states \cite{ss1} and their
topological classification \cite{ss2}, generation of topologically
non-trivial quantum states \cite{topo1}, several types of dynamical
transitions \cite{dt1,dt2}, possibility of tuning ergodicity
properties of driven systems \cite{ra1}, dynamical localization
\cite{rev3,dl1,dl2} and dynamical freezing \cite{df1,df2}.

The properties driven systems are encoded in their evolution
operator
\begin{eqnarray}
U(t,0)= {\mathcal T}_t \exp\left[-\frac{i}{\hbar} \int_0^t H(t') dt'
\right], \label{udef1}
\end{eqnarray}
where $H(t)$ denotes the Hamiltonian of the system, and ${\mathcal
T}_t$ denotes time ordering. This operator maps the initial state of
a driven system at $t=0$ to its final state at time $t$:
$|\psi(t)\rangle = U(t,0) |\psi(0)\rangle$. For periodically driven
systems characterized by a period $T=2 \pi/\omega_D$, where
$\omega_D$ is the drive frequency, the evolution operator for all
times $t_0=n_0T$ (where $n_0 \in Z$ is the number of drive periods)
is given in terms of the Floquet Hamiltonian $H_F$ by $U(t_0,0) =
\exp[-i H_F n_0 T/\hbar]$ \cite{floqref}. This form of $U$ is a
consequence of time periodicity of the driven system and is
independent of system details. It is well-known that all information
about the stroboscopic time evolution of the system is encoded in
$H_F$ \cite{rev3}. Moreover the eigenfunctions of $H_F$ provides one
with information regarding the long-time steady states of such
driven systems \cite{rev4,rigol1}.

The computation of the Floquet Hamiltonian in such driven system
poses a significant challenge. In the high-drive frequency regime,
one can resort to systematic Magnus expansion and compute the
Floquet Hamiltonian \cite{magnusref1}. Several forms of these
expansion have been used in the literature
\cite{magnusref2,magnusref3}. However, all of them invariably fails
at intermediate and low drive frequencies (when the drive frequency
approximately equals system energy scales); moreover, estimating the
radius of convergence of such expansion poses significant
theoretical challenge \cite{saitoref1}. For discrete drive protocols
(such as periodic kicks or square pulse protocols), it seems
possible to provide a resummation of such Magnus series using
replica trick \cite{anatoli1}; however, this procedure can not be
carried out for continuous drive protocols in a straightforward
manner. Another technique which has been used for computing $H_F$ in
such driven systems is the flow equation method which provided
significantly better results than Magnus expansion at intermediate
frequencies \cite{feref1}; however the stability of fixed points
obtained by this method seems difficult to asses for interacting
systems in the low frequency regime. For low or intermediate drive
frequencies, analytic computation of Floquet Hamiltonian thus seems
to be more difficult. For a class of integrable models, an
adiabatic-impulse approximation has been used to compute $H_F$
\cite{adimpref1}. However, such approximations have no obvious
generalization for non-integrable interacting systems. More
recently, a Floquet perturbation theory has attempted to put the
high- and the low-frequency approximations to $H_F$ at the same
footing; such a theory has been applied to a class of integrable
models and is shown to produce accurate description of $H_F$
\cite{babak1}. However, its application to interacting Hamiltonians
remains an unsolved problem.

The numerical computation of the eigenspectra of $H_F$ for
interacting non-integrable has also been attempted in several works
\cite{numerics1}. Typically such procedure is simple for piecewise
continuous drive protocols; for these, the time ordering ${\mathcal
T}_t$ can be easily done. For example, for a square pulse protocol
for which $H(t)= H_a$ for $0 \le t\le T/2$ and $H_b$ for $T/2 < t
\le T$, one has
\begin{eqnarray}
U(T,0) = \exp[-i H_b T/(2\hbar)] \exp[-i H_a T/(2 \hbar)].
\label{uevol2}
\end{eqnarray}
Consequently, $U$ and hence $H_F$ can be computed from the knowledge
of eigenstates and eigenvalues of $H_a$ and $H_b$. In contrast, for
continuous drive, one typically needs to evaluate $U$ by
constructing Trotter product of $U_i = U(t_{i-1}+\delta
t_i,t_{i-1})$ computed for infinitesimal time slices $\delta t_i$:
$U(T,0)= U_1 U_2 ... U_N$ with $T= N \delta t_i$. The width, $\delta
t_i = T/N$ of these slices depends on energy scales of the problem
and the rate at which $H(t)$ changes. Such trotterization of $U$ is
clearly computationally intensive and can not be reliably done for
interacting systems for large system size. Thus numerical studies of
periodically driven systems has been mostly carried out with
piecewise continuous protocol.

In this work we apply a Floquet perturbation theory (FPT) on a
continually driven interacting Fermi systems in the weak interaction
limit. The Floquet Hamiltonian so obtained can be used to study
dynamics of such fermions in arbitrary dimensions; in this work, we
shall apply them to interacting fermions chains. The non-interacting
fermion chains has been studied in several context
\cite{antal1,antal2,eisler1}; however, aspects of dynamics of the
interacting chain has only been recently addressed for a piecewise
continuous drive protocol \cite{dl2}. For such chains, the relevant
energy scales are given by ${\mathcal J}_0$ which is the amplitude
of the kinetic term, $V_0$ which is the interaction strength, and
$\hbar \omega_D$ which is the energy scale coming from the drive. We
develop the FPT for $ V_0 \ll {\mathcal J}_0$; our results indicate
that there exists a wide frequency range $ V_0 \le \hbar \omega_D$
where such a FPT provides accurate information about the system
dynamics. This feature needs to be contrasted with the Magnus
expansion which typically works for $\hbar \omega_D \ge {\mathcal
J}_0$. We note here that such FPT has been discussed for spin
systems subjected to piecewise continuous drive protocols earlier
\cite{fpt1,fpt2,ra1,df2} and in context of Floquet scattering theory
\cite{tb1}. Here we shall use the formalism developed in Ref.\
\onlinecite{tb1} to addresses the dynamics of the continually driven
fermion chain.

The central results that we obtain from this study are as follows.
First, we provide an semi-analytic expression for $H_F$ of the
driven interacting fermions and compare it to its counterpart
obtained from Magnus expansion for a fermionic chain. To this end,
we use eigenspectra of $H_F$ to compute the state
$|\psi(T)\rangle_{\rm pert}$ of the driven chain after one drive
cycle starting from a random initial state. We compute its overlap
$F$ with $|\psi(T)\rangle_{\rm exact}$ computed using exact
diagonalization (ED) starting from the same initial state. We find
that for any random initial state and for all drive frequencies $V_0
\le \hbar \omega_D \le {\mathcal J}_0$, $F$, computed using FPT, has
a much higher value than its counterpart obtained using the Magnus
expansion. We chart out the variation of $F$ with both $\omega_D$
and $V_0$ and thus delineate the regime of validity of FPT for the
system. Second, we discuss the approach of the system to its steady
state via computation of the expectation value of $H_{\rm av} =
\int_0^T H(t) dt/T$ in the steady state. We express the steady state
expectation value of $H_{\rm av}$ using a dimensional quantity $Q$
which is a bounded function assuming values between $0$ and $-1$
\cite{rigol1}. The construction of $Q$ is designed so that $Q=0$
when $\langle H_{\rm av}\rangle_{\rm steady state}$ assumes the
infinite temperature steady state value as predicted by eigenstate
thermalization hypothesis (ETH); in contrast $Q=-1$ when the steady
state is same as the initial state \cite{rigol1}. We find using FPT
that $Q$, for finite driven chains, lies between these two values
signifying the presence of sub- or super-thermal steady states for a
wide range of drive frequencies. We relate such behavior to the
structure of the Floquet eigenspectrum of the system. We also
compute the Shannon entropy of the driven system using its Floquet
spectrum obtained from our FPT analysis and show that it can serve
as an qualitative indicator of localization-delocalization crossover
in these driven finite chains. Third, we study the crossover of
localized to delocalized behavior of fermions in the driven system.
It is well-known that the non-interacting fermion model exhibit
perfect dynamical localization for continual drive protocol used in
this work; here, we study the fate of this localization for a driven
finite fermion chain in the steady state as a function of drive
frequency. For finite chains, we find the existence of a crossover
between localized and delocalized steady states at intermediate
frequencies $\hbar \omega_D \sim {\mathcal J}_0 /2 \gg V_0$. We
discuss the implication of such a crossover for large chains in the
thermodynamic limit and discuss experiments which can test our
theory.

The rest of the paper is organized as follows. In Sec.\ \ref{fpt},
we derive the Floquet Hamiltonian using FPT and compute the fidelity
between wavefunctions after a drive cycle obtained from it and that
obtained from exact numerics. This is followed by Sec.\ \ref{ssapp},
where we compute $Q$ and the Shannon entropies for finite sized
interacting fermion chains using both ED and the eigenspectrum of
$H_F$ obtained via FPT. Next, in Sec.\ \ref{dynloc}, we study
dynamical localization in such fermionic chains and compare results
obtained from ED and the Floquet Hamiltonian over a range of drive
frequencies and interaction strengths. Finally, in Sec.\ \ref{diss},
we summarize our results, discuss experiments which can test them,
and conclude.

\section{Floquet Hamiltonian}
\label{fpt}

In this section, we shall use the Floquet perturbation theory
developed in Ref.\ \onlinecite{tb1} and apply it to weakly
interacting spinless fermions. Our analysis will be applicable for
fermions in arbitrary dimensions; however, all numerical studies
shall be restricted to 1D fermion chains.

The Hamiltonian for such a fermionic system is given by
$H(t)=H_0(t)+H_1$, where
\begin{eqnarray}
H_0(t) &=& {\mathcal J}(t) \sum_{\vec k} \epsilon_{\vec k} c_{\vec
k}^{\dagger}
c_{\vec k}  \nonumber\\
H_1 &=& \sum_{\vec k_1, \vec k_2, \vec q} V_q c_{\vec k_1}^{\dagger}
c_{\vec k_2}^{\dagger} c_{\vec k_2 -\vec q} c_{\vec k_1 +\vec q}
\label{fermham1}
\end{eqnarray}
where ${\mathcal J}(t)= {\mathcal J}_0 f(t)$ is the time dependent
amplitude of the kinetic term for the fermions, $c_{\vec k}$ denotes
fermion annihilation operator, and $f(t)$ species the drive
protocol. In this work, we shall choose $f(t) =\cos(\omega_D t)$
where $\omega_D$ is the drive frequency. Moreover, in what follows,
we shall use $V_q = \sum_{i=1,z} V_0 \exp[i q_i a_i]$, where $\vec
a$ denotes the lattice spacing between two neighboring fermions and
$z$ is the coordination number of the lattice with $z=2d$ for a
hypercubic lattice in $d$ dimension. This choice is made so that
$H_1$ is the Fourier transform of $H'_1= V_0 \sum_{\langle \vec j_1
\vec j_2\rangle} \hat n_{\vec j_1} \hat n_{\vec j_2}$, where
$\langle \vec j_1 \vec j_2\rangle$ implies that $\vec j_1$ and $\vec
j_2$ are neighboring sites and $\hat n_{\vec j} = c_{\vec
j}^{\dagger} c_{\vec j}$ is the fermion density operator; $H_1$, for
$V_0>0$, thus represents fermions with nearest neighbor repulsive
interaction. Here $\epsilon_k$ denotes the fermion dispersion in
momentum space; for fermions with nearest neighbor hopping on a
$d$-dimensional hypercubic lattice $\epsilon_{\vec k} = -
\sum_{i=1,d} \cos(k_i a_i)$.

For $V_0=0$, the evolution operator $U_0(t,0)$ for the
non-interacting Hamiltonian can be easily constructed. This is
given, for $f(t) = \cos \omega_D t$, by
\begin{eqnarray}
U_0(t,0) &=& \exp \left[-i \frac{{\mathcal J}_0}{\omega_D}
\sin(\omega_D t) \sum_{\vec k} \epsilon_{\vec k} \hat n_{\vec k}
\right] \label{evolzero}
\end{eqnarray}
where $\hat n_{\vec k} =c_{\vec k}^{\dagger} c_{\vec k}$ and here,
and in the rest of this work, we set $\hbar$ to unity unless
mentioned otherwise. We note that $U_0(t,0)$ is diagonal in the
number basis at all times, and that $U_0(T,0)=1$, so that $H_F^{(0)}
= 0$ for the non-interacting fermions. This in turn implies that
such fermions do not show stroboscopic evolution and the
wavefunction after $n_0 \in Z$ drive cycles satisfies
$|\psi(n_0T)\rangle = |\psi_0 \rangle$ for any initial wavefunction
$|\psi_0\rangle$.

The first non-trivial term in the Floquet Hamiltonian can be
perturbatively computed using standard time dependent perturbation
theory. One gets, for first order correction to the evolution
operator $U(T,0)$ denoted by $U_1(T,0)$,
\begin{eqnarray}
U_1(T,0) &=&  -i \int_0^T H_1^{I}(t) dt \label{uevol1}
\end{eqnarray}
where $H_1^{I}=U_0^{\dagger}(t,0) H_1 U_0(t,0)$ denotes the
interacting part of $H$ in the interaction picture. To obtain the
Floquet Hamiltonian from here, we first compute the matrix element
of $U_1$ between two arbitrary many-body number states
$|\alpha\rangle = |n_{\vec k_1}^{\alpha} .... n_{\vec
k_n}^{\alpha}\rangle $ and $|\beta\rangle = |n_{\vec k_1}^{\beta}
.... n_{\vec k_n}^{\beta}\rangle$. A straightforward calculation
yields
\begin{widetext}
\begin{eqnarray}
\langle \alpha|U_1(T,0)|\beta \rangle &=& -i \sum_{\vec k_1,\vec
k_2, \vec q} \int_0^T dt e^{i \frac{{\mathcal J}_0}{\omega_D} \,
\mu_{\vec k_1 \vec k_2 \vec q}^{\alpha \beta} \,\sin(\omega_D t)
}V_{\vec q} \, \Gamma^{\alpha \beta}_{\vec k_1 \vec k_2 \vec q}
\nonumber\\
\mu_{\vec k_1 \vec k_2 \vec q}^{\alpha \beta} &=& \sum_{\vec k' =
\vec k_1, \vec k_2, \vec k_2 -\vec q, \vec k_1+\vec q}
\epsilon_{\vec k'} (n_{\vec k'}^{\alpha} - n_{\vec k'}^{\beta}),
\quad \Gamma^{\alpha \beta}_{\vec k_1 \vec k_2 \vec q} = \langle
\alpha |c_{\vec k_1}^{\dagger} c_{\vec k_2}^{\dagger} c_{\vec k_2
-\vec q} c_{\vec k_1 +\vec q} |\beta\rangle \label{ufirstorder1}
\end{eqnarray}
\end{widetext}
The matrix elements $\Gamma^{\alpha \beta}_{\vec k_1 \vec k_2 \vec
q} $ play a central role in determining $H_F$ and can be written as
\begin{widetext}
\begin{eqnarray}
\Gamma^{\alpha \beta}_{\vec k_1 \vec k_2 \vec q} &=& (-1)^{f_{\vec
k_1 \vec k_2 \vec q}^{\alpha \beta}} \delta_{n_{\vec k_1}^{\alpha},
n_{\vec k_1}^{\beta}+1} \delta_{n_{\vec k_2}^{\alpha}, n_{\vec
k_2}^{\beta}+1} \delta_{n_{\vec k_2-\vec q}^{\alpha}, n_{\vec k_2
-\vec q}^{\beta}-1} \delta_{n_{\vec k_1 +\vec q}^{\alpha}, n_{\vec
k_1 +\vec q}^{\beta}-1}, \quad {\rm for} \,\, \vec q \ne 0 \, \,{\rm
and
} \,\, \vec k_2 -\vec k_1 \ne \vec q \nonumber\\
&=& \delta_{\alpha \beta} n_{\vec k_1}^{\alpha} n_{\vec
k_2}^{\alpha} (\delta_{\vec q,0} -\delta_{\vec k_2, \vec k_1 +\vec
q}), \quad {\rm otherwise} \label{matelexp}\\
f_{\vec k_1 \vec k_2 \vec q}^{\alpha \beta} &=& \sum_{\vec
k=0}^{\vec k_1} n_{\vec k}^{\alpha} +  \sum_{\vec k=0}^{\vec k_2}
n_{\vec k}^{\alpha'} +\sum_{\vec k=0}^{\vec k_2-\vec q} n_{\vec
k}^{\beta'} +\sum_{\vec k=0}^{\vec k_1+\vec q} n_{\vec k}^{\beta},
\quad \langle\alpha'| = \langle \alpha| c_{\vec k_1}^{\dagger} \,\,
{\rm and}\, \, |\beta'\rangle = c_{\vec k_1+\vec q} |\beta \rangle
\nonumber
\end{eqnarray}
\end{widetext}
Using the identity $\exp[i a_0 \sin(\omega_D t)]= \sum_{n =
-\infty}^{\infty} J_n(a_0) \exp[i n \omega_D t]$, it is easy to
evaluate the integral in Eq.\ \ref{ufirstorder1}. This yields
\begin{eqnarray}
\langle \alpha|U_1(T,0)|\beta \rangle  &=& -i \sum_{\vec k_1,\vec
k_2, \vec q} V_{\vec q} T J_0\left[\frac{{\mathcal J}_0 \mu_{\vec
k_1 \vec k_2 \vec q}^{\alpha \beta}}{\omega_D} \right] \Gamma_{\vec
k_1 \vec k_2 \vec q}^{\alpha \beta} \label{ufirstorder2}
\end{eqnarray}
Since $H_0^F=0$ and at this order $U_1(T,0) \simeq  1 - i H^{(1)}_F
T$, one can read off the expression for the matrix element of the
first order Floquet Hamiltonian $H_F^{(1)}$ to be \cite{tb1}
\begin{eqnarray}
\langle \alpha|H_F^{(1)}|\beta \rangle  &=& \sum_{\vec k_1,\vec k_2,
\vec q} V_{\vec q}  J_0\left[\frac{{\mathcal J}_0 \mu_{\vec k_1 \vec
k_2 \vec q}^{\alpha \beta}}{\omega_D} \right] \Gamma_{\vec k_1 \vec
k_2 \vec q}^{\alpha \beta} \label{fham1}
\end{eqnarray}
We note that for large $\omega_D \gg {\mathcal J}_0, V_0$,
$J_0[{\mathcal J}_0 \mu_{\vec k_1 \vec k_2 \vec q}^{\alpha
\beta}/\omega_D ] \to 1$. In this limit, Eq.\ \ref{fham1} yields the
Magnus result: $H_F^{(1)\, {\rm magnus}} = H_1$, where we have used
Eq.\ \ref{ufirstorder1} to represent $\Gamma^{\alpha \beta}_{\vec
k_1 \vec k_2 \vec q}$ in terms of fermion creation and annihilation
operators. However, for $\omega_D \sim {\mathcal J}_0$ such
simplification does not occur and Eq. \ref{fham1} predicts a much
more complicated structure for $H_F^{(1)}$. As we shall see, this
deviation from the Magnus result is key to an accurate description
of the system at intermediate frequencies. We note here that the
matrix elements of $H_F^{(1)}$ are significant when
$|E_{\alpha}-E_{\beta}| \sim {\rm O}(\omega_D)$; for states
$|\alpha\rangle$ and $|\beta \rangle$ with larger energy difference,
$J_0[ {\mathcal J}_0 \mu^{\alpha \beta}/\omega_D] \sim
[\omega_D/({\mathcal J}_0 \mu^{\alpha \beta})]^{1/2} \to 0$ leading
to small matrix elements of $H_F^{(1)}$ between such states.

Next, we compute the second order term in the Floquet Hamiltonian.
To this end, we note that the second order correction to $U(T,0)$ is
given by
\begin{eqnarray}
U_2(T,0) &=& (-i)^2 \int_0^T dt_1 H_F^I(t_1) \int_0^{t_1} dt_2
H_F^I(t_2) \label{uevolsecondorder1}
\end{eqnarray}
Substituting an intermediate many-body number state $|\gamma\rangle
=|n_{\vec k_1}^{\gamma} ...n_{\vec k_N}^{\gamma}\rangle$, one
obtains after a straightforward calculation
\begin{widetext}
\begin{eqnarray}
\langle \alpha| U_2(T,0) |\beta\rangle &=& (-i)^2 \sum_{\gamma}
\sum_{\vec k_1, \vec k_2, \vec q}\sum_{\vec k'_1, \vec k'_2, \vec
q'} \sum_{m,n=-\infty}^{\infty} V_{\vec q} V_{\vec q'}
J_n\left[\frac{{\mathcal J}_0 \mu_{\vec k_1 \vec k_2 \vec q}^{\alpha
\gamma}}{\omega_D} \right] J_m\left[\frac{{\mathcal J}_0 \mu_{\vec
k'_1 \vec k'_2 \vec q'}^{\gamma \beta}}{\omega_D} \right]
\Gamma_{\vec k_1 \vec k_2 \vec q}^{\alpha \gamma} \Gamma_{\vec k'_1
\vec k'_2 \vec q'}^{\gamma \beta} S_{nm}(T) \nonumber\\
S_{nm}(T) &=& \int_0^T e^{i n \omega_D t_1} dt_1 \int_0^{t_1} e^{i m
\omega_D t_2} dt_2  \nonumber\\
&=& \frac{T}{i \omega_D} \left[ (1-\delta_{n0})(1-\delta_{m0})
\frac{\delta_{n-m}}{m} +(1-\delta_{n 0})\frac{\delta_{m0}}{n} -
(1-\delta_{m 0})\frac{\delta_{n0}}{m} \right] + \delta_{n0}
\delta_{m0} T^2/2 \label{u2exp}
\end{eqnarray}
\end{widetext}
We note that the least term in Eq.\ \ref{u2exp} leads to a term in
$U_2(T,0)$ which is identical to $U_1^2(T,0)/2$. Using this
observation one can read off the expression for the matrix elements
of the second order term in the Floquet Hamiltonian as
\begin{widetext}
\begin{eqnarray}
\langle \alpha |H_F^{(2)}|\beta\rangle &=&  \sum_{\gamma} \sum_{\vec
k_1,\vec k_2, \vec q} \sum_{\vec k'_1,\vec k'_2 \vec q'} \,
\sum_{n=1}^{\infty} \frac{2 V_{\vec q} V_{\vec q'}}{(2n+1) \omega_D}
\left[J_{2n+1}\left[\frac{{\mathcal J}_0}{\omega_D} \mu_{\vec k_1
\vec k_2 \vec q}^{\alpha \gamma} \right] J_0\left[\frac{{\mathcal
J}_0}{\omega_D} \mu_{\vec k'_1 \vec k'_2 \vec q'}^{\gamma \beta}
\right] \right.\nonumber\\
&& \left. - J_0\left[\frac{{\mathcal J}_0}{\omega_D} \mu_{\vec k_1
\vec k_2 \vec q}^{\alpha \gamma} \right]
J_{2n+1}\left[\frac{{\mathcal J}_0}{\omega_D} \mu_{\vec k'_1 \vec
k'_2 \vec q'}^{\gamma \beta} \right] \right] \Gamma_{\vec k_1 \vec
k_2 \vec q}^{\alpha \gamma} \Gamma_{\vec k'_1 \vec k'_2 \vec
q'}^{\gamma \beta} \label{fham2}
\end{eqnarray}
\end{widetext}
where we have used the identity $J_n(x)= (-1)^n J_{-n}(x)$.

Eqs.\ \ref{fham1} and \ref{fham2} yield the matrix elements of the
Floquet Hamiltonian for weakly interacting fermions. We note the
following features about these equations. First, we find that
$H_F^{(2)} \to 0$ for $\omega_D \to \infty$; thus our result
reproduces the fact that the Floquet Hamiltonian, as obtained from
Magnus expansion, does not have any finite second order term:
$H_F^{(2){\rm magnus}}=0$. This can be easily checked from a
straightforward direct calculation. Second, we note that the second
order matrix elements involves a sum over virtual many-body state
$\gamma$; thus $H_F^{(2)}$, in contrast to its first order
counterpart, may have finite contribution for $|E_{\gamma}
-E_{\alpha}|, |E_{\gamma}-E_{\beta}| \gg \omega_D$. Third, an
extension of these results to higher order perturbation theory is
straightforward although the results become quite cumbersome. But
quite generally, it is easy to see that the $p^{\rm th}$ order term
in the perturbation expansion for $H_F$ contains $H_F^{(p)} \sim
V_0^p /(\omega_D)^{p-1} J_{n_1}(x_1) ... J_{n_p}(x_p)$, where
$n_1\,...\,n_p$ are integers and $x_j \sim {\mathcal J}_0/\omega_D$.
Thus for small enough $\omega_d$ where $x_j \gg 1$, $J_n(x_j) \sim
(x_j)^{-1/2}$, one has $H_F^{(p)} \sim V_0^p/(\omega_D^{p/2-1})$.
This implies that for terms where all $x_j$ s are large,  the
perturbation theory will surely breakdown around $V_0 \sim
\sqrt{\omega_D}$ for large $p$. In practise not all $J_{n_j}$ s need
to have large arguments simultaneously and one therefore expects the
perturbation theory to break down at higher $\omega_D \le V_0$.
Numerically we find that the perturbation theory stars deviating
from the exact result around $V_0 \simeq \omega_D$. Thus FPT is
expected to provide accurate results for $\omega_D \ge V_0$.
Finally, the matrix elements of both $H_F^{(1)}$ and $H_F^{(2)}$
constitute results which can not be obtained using perturbation in
$1/\omega_D$; thus they constitute resummation of all ${\rm
O}(V_0/{\mathcal J}_0)$ and ${\rm O}(V_0^2/{\mathcal J}_0^2)$ terms
of the Magnus expansion. The existence of such a resummed Floquet
Hamiltonian is one of the main results of this work.

\begin{figure}
\rotatebox{0}{\includegraphics*[width=0.49\linewidth]{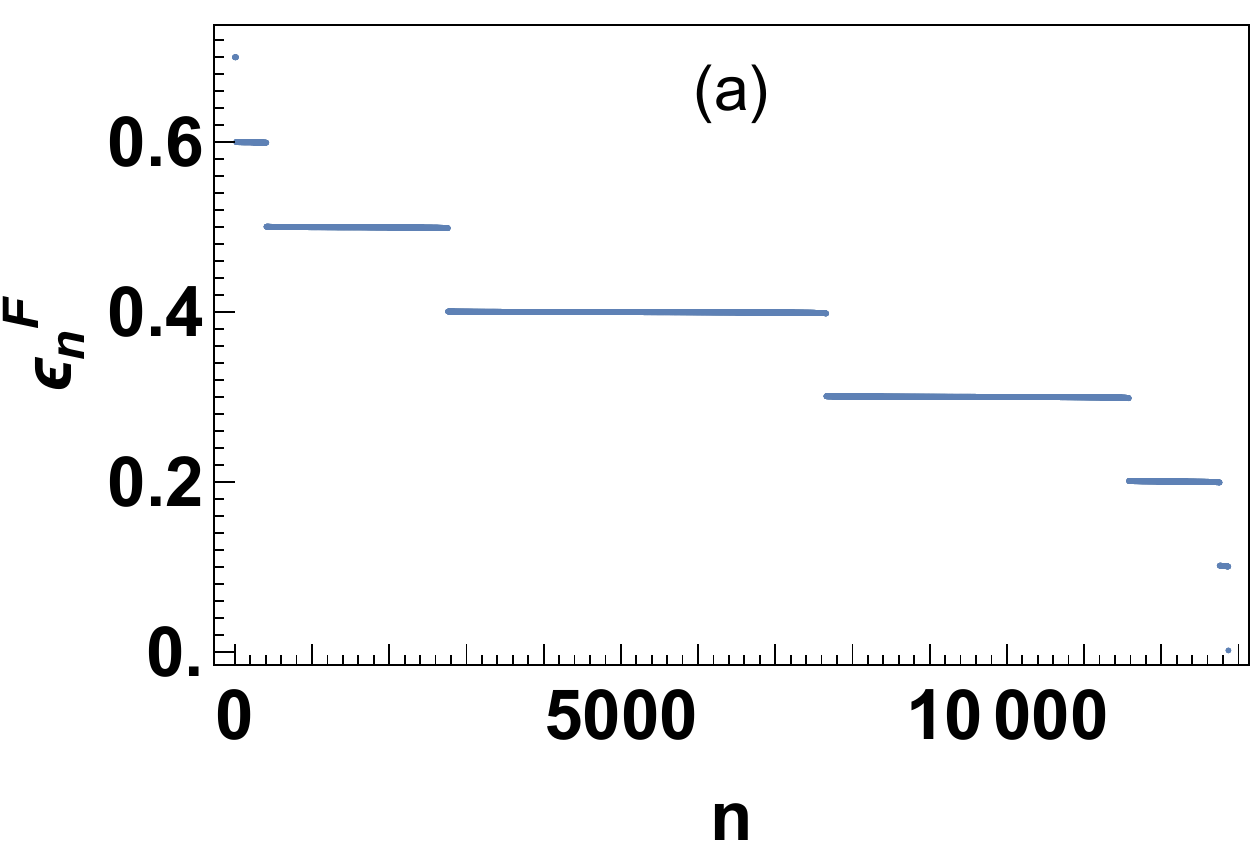}}
\rotatebox{0}{\includegraphics*[width=0.49\linewidth]{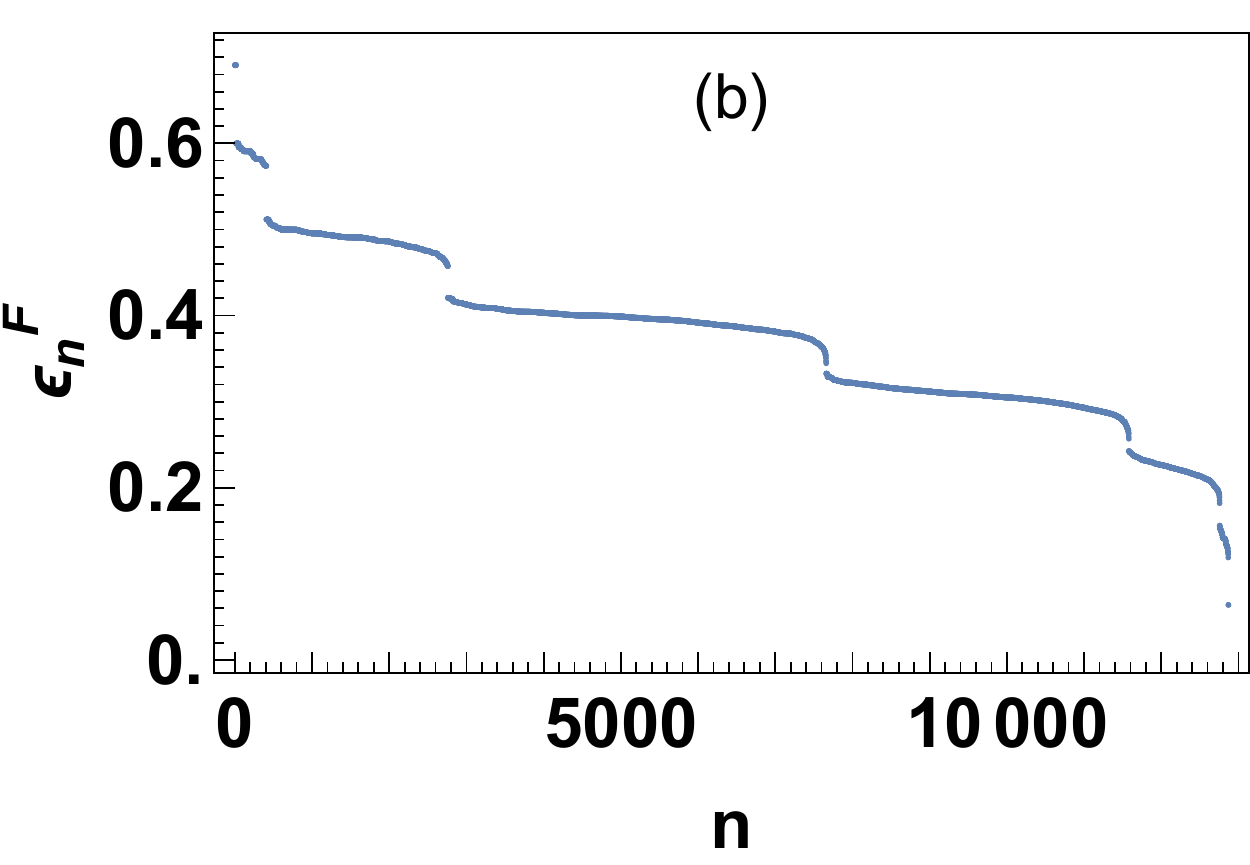}} \\
\rotatebox{0}{\includegraphics*[width=0.49\linewidth]{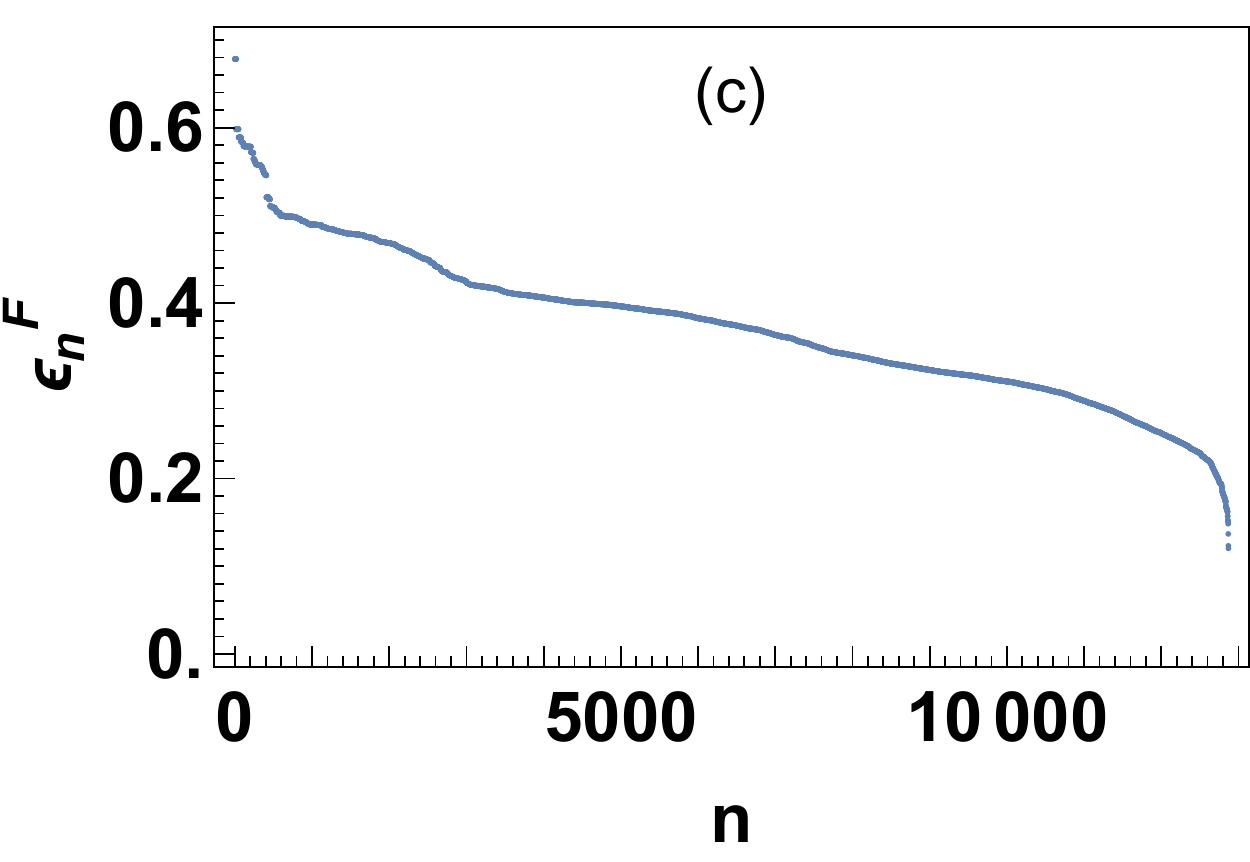}}
\rotatebox{0}{\includegraphics*[width=0.49\linewidth]{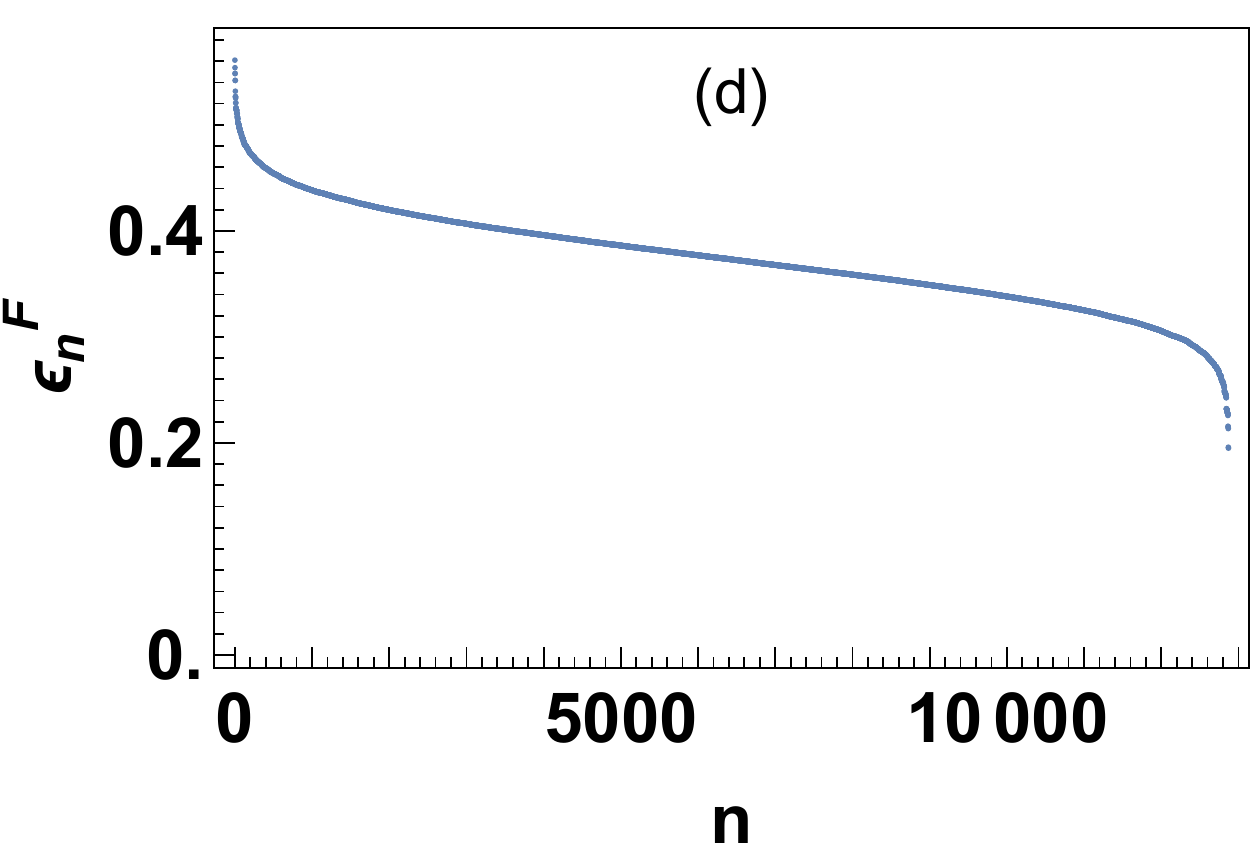}}
\caption{Plot of the Floquet eigenvalues of a 1D interacting fermion
chain as a function of the quantum number $n$ for $V_0=0.1$ and (a)
$\omega_D=10$, (b) $\omega_D=1.6$, (c) $\omega_D=1$ and (d)
$\omega_D=0.1$. The Floquet spectrum displays flat bands at high
$\omega_D/V_0$. For all plots all energies and frequencies are
measured in units of ${\mathcal J}_0$, $\hbar$ is set to unity, and
the chain length is $L=16$. See text for details.} \label{fig1}
\end{figure}

In the remaining part of this section, we shall compare these
results with exact numerical result using ED for 1D fermionic chain.
To this end, we first diagonalize the perturbative Floquet
Hamiltonian whose matrix elements are given by $H_{F}^{(1)} +
H_F^{(2)}$ (Eqs.\ \ref{fham1} and \ref{fham2}) by using exact
diagonalization for finite sized chains with $L \le 16$. We denote
these eigenvalues as $\epsilon_n^F$; the corresponding eigenvectors
are given by $|\chi_n\rangle$. These eigenvalues are plotted in
Fig.\ \ref{fig1} as a function of their index $n$ for several
representative values of $\omega_D/{\mathcal J}_0$ and
$V_0/{\mathcal J}_0=0.1$. We note that the spectrum display flat
band structure at $\omega_D \gg V_0, {\mathcal J}_0$; in contrast,
it starts to show dispersing behavior for $\omega_D \simeq {\mathcal
J}_0$. This difference between the high frequency and low-frequency
behavior can be understood as follows. For the non-interacting
Hamiltonian ($H=H_0$), the Floquet spectrum displays a perfect flat
band at zero quasienergy (since $H_F^{(0)}=0$). At high-frequencies
$\omega_D \gg {\mathcal J}_0$, where $H_F \simeq H_1$, the
interaction partially lifts this degeneracy and the eigenspectra
shows multiple flat bands. Upon further decreasing $\omega_D$, these
bands start to disperse; this behavior is first seen around
$\omega_D /{J}_0 \sim 1$ where the Bessel functions in Eqs.\
\ref{fham1} and \ref{fham2} starts to deviate from their values for
$\omega_D \gg {\mathcal J}_0$. Also around these frequencies,
$H_F^{(2)}$ starts to contribute significantly to $H_F$. Finally,
when $\omega_D \sim V_0 \ll {\mathcal J}_0$, the Floquet bands
become completely dispersive in nature. We note that in contrast,
$H_F^{\rm magnus}=H_1$ always shows flat bands similar to Fig.\
\ref{fig1}(a); it does not capture the evolution of the band
dispersion with $\omega_D$.

To compare between the perturbative analytic approach and exact
numerics, we compare the wavefunction overlap $F$ between
wavefunction $|\psi(T)\rangle_{\rm pert}$ obtained using FPT and
$|\psi(T)\rangle_{\rm exact}$ computed using exact numerical
solution. As discussed earlier, computation of eigenstectra of
$U(T,0)$ exactly is an extremely computationally intensive procedure
with such a continuous drive. Hence we use this method to show the
accuracy of the FPT approach.

To this end, we first rewrite the evolution operator in terms of the
Floquet quasienergies $\epsilon_n^F$ and eigenfucntions
$|\chi_n\rangle$ as
\begin{eqnarray}
U_{\rm pert}(T,0) &=& \sum_n e^{-i \epsilon_n^F T}
|\chi_n\rangle\langle \chi_n|  \label{uevolpert1}
\end{eqnarray}
This allows us to write, for an arbitrary initial state
$|\psi_0\rangle$, the state after one drive cycle as
\begin{eqnarray}
|\psi(T)\rangle_{\rm pert} &=& \sum_n c_n e^{-i \epsilon_n^F T}
|\chi_n \rangle, \quad c_n = \langle \chi_n|\psi_0\rangle
\label{wavpert}
\end{eqnarray}

Next, we obtain $|\psi(T)\rangle_{\rm exact}$ as follows. We first
use ED to obtain eigenvalues $\epsilon_n$ and eigenfunctions
$|\phi_n\rangle$ for the fermionic Hamiltonian given by Eq.\
\ref{fermham1} at $t=0$. In terms of these exact eigenstates one can
write the starting state $|\psi_0\rangle = \sum_n d_n^{(0)}
|\phi_n\rangle$. Since $|\phi_n\rangle$ forms a complete basis, the
wavefunction $|\psi(t)\rangle_{\rm exact}$ for any $t$ can be
expressed as $|\psi(t)\rangle_{\rm exact} = \sum_n d_n(t) e^{-i
\epsilon_n t} |\phi_n\rangle$ where
\begin{eqnarray}
i \partial_t d_n(t) &=& \sum_{mn} \eta_{nm}(t)
d_m(t)\nonumber\\
\eta_{mn}(t) &=& \langle \phi_n | H_0(t)-H_0(0)| \phi_m\rangle, \,\,
d_n(0)=d_n^{(0)}  \label{wavexact}
\end{eqnarray}
We solve Eq.\ \ref{wavexact} numerically to obtain
$|\psi(T)\rangle_{\rm exact}$.

\begin{figure}
\rotatebox{0}{\includegraphics*[width=0.49\linewidth]{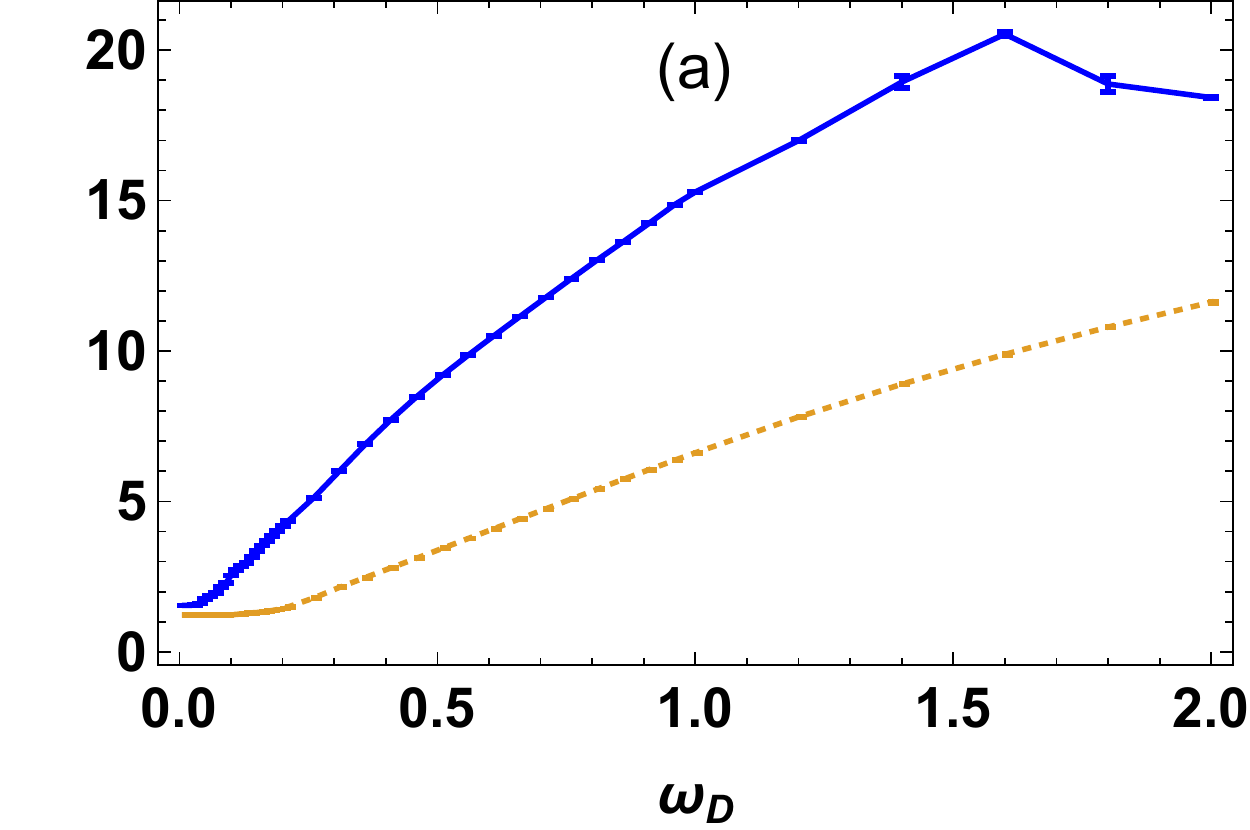}}
\rotatebox{0}{\includegraphics*[width=0.49\linewidth]{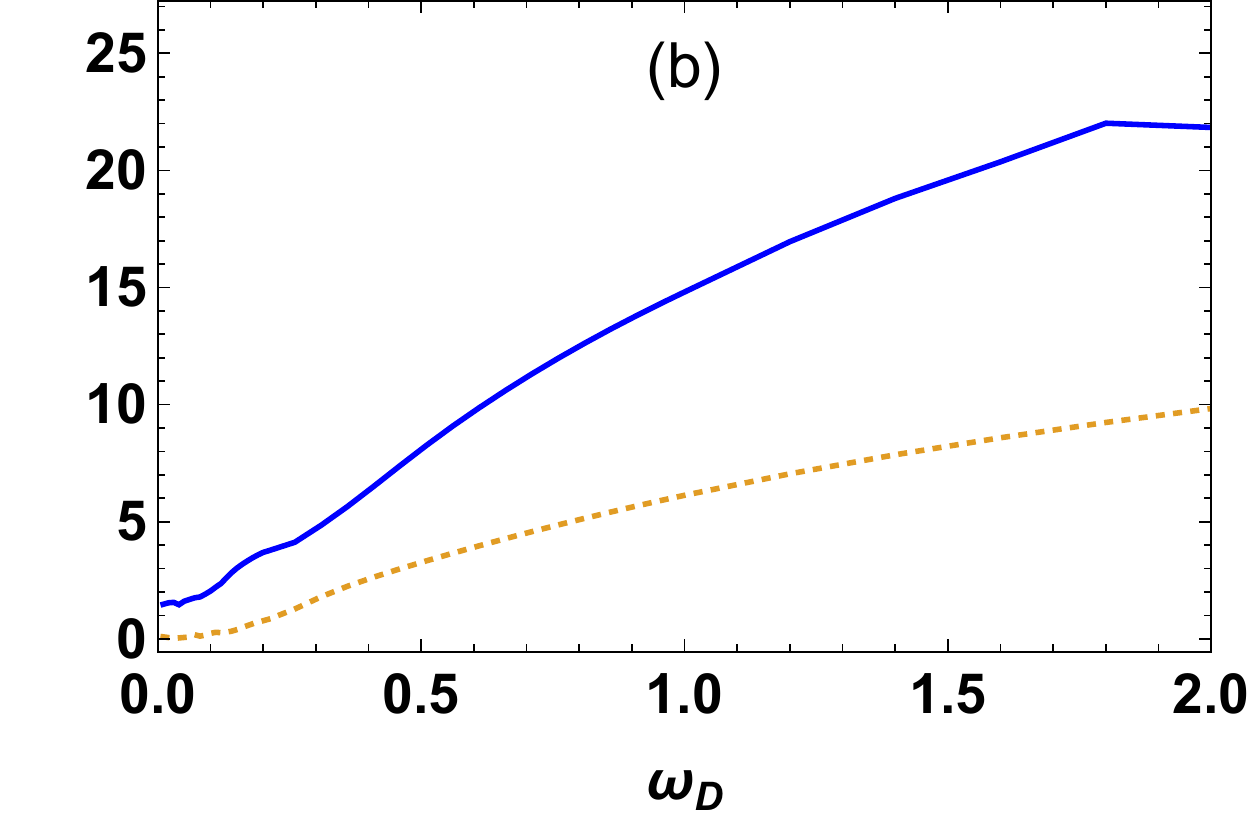}} \\
\rotatebox{0}{\includegraphics*[width=0.49\linewidth]{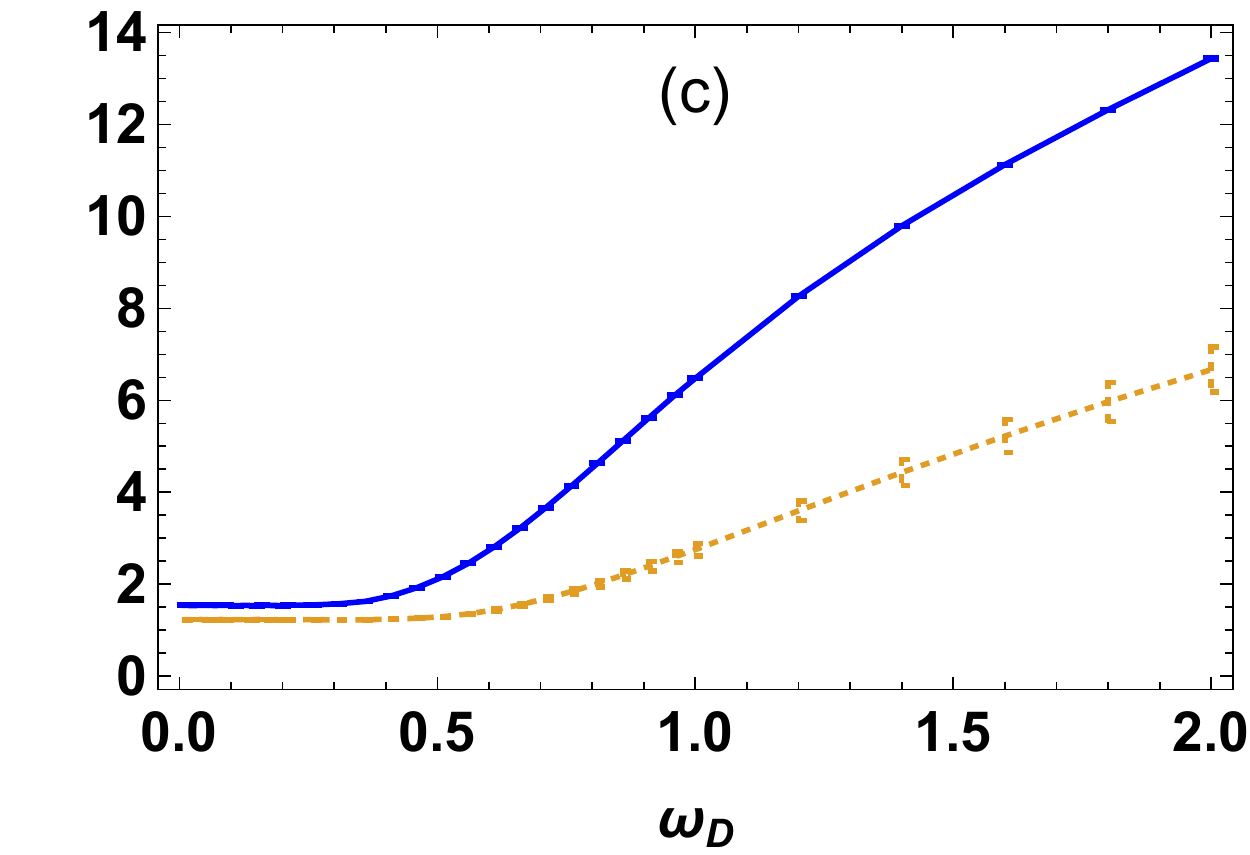}}
\rotatebox{0}{\includegraphics*[width=0.49\linewidth]{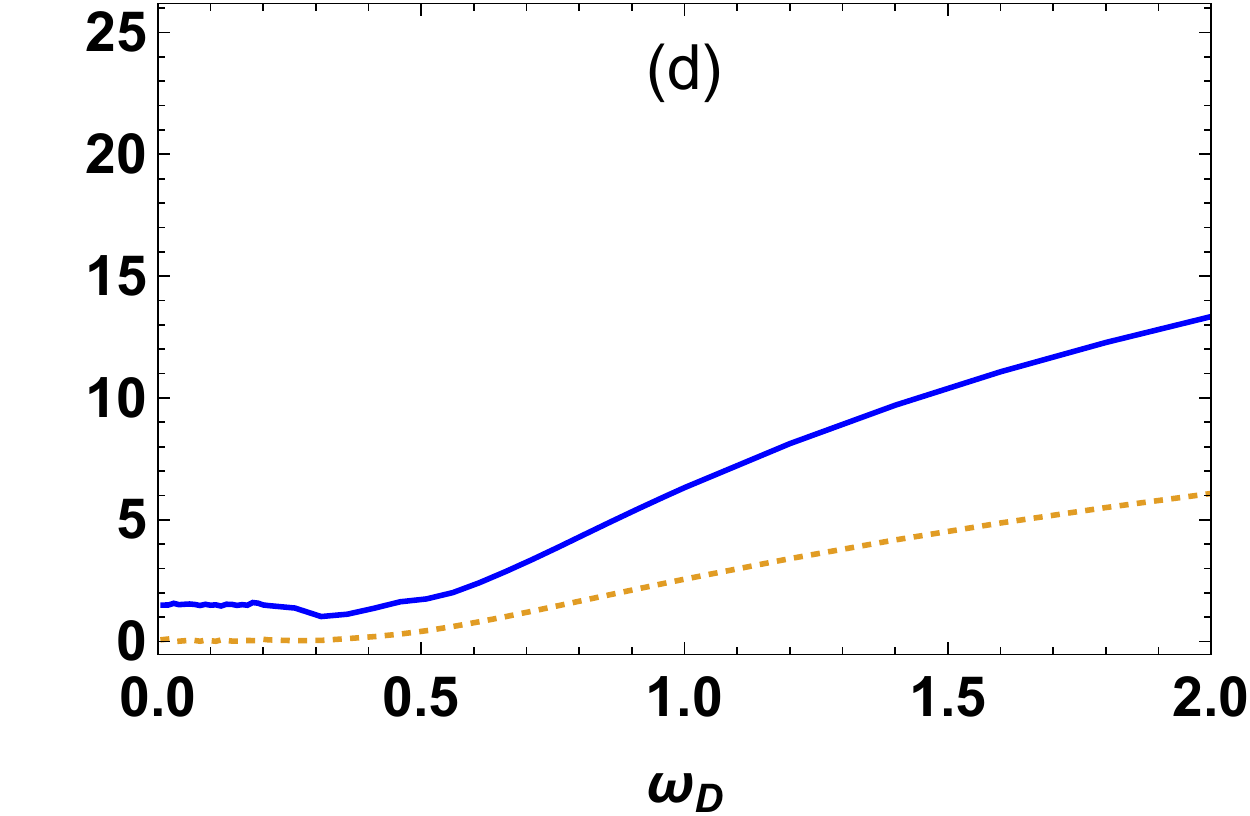}}
\caption{(a) Plot of $C_{\rm av}$ (blue solid line) and $C_{\rm
av}^m$ (yellow dotted line) for $V_0=0.1$ as a function of
$\omega_D$. (b) Plot of $C_{\rm prod}$ (blue solid line) and $C_{\rm
prod}^m$ (yellow dotted line) as a function of $\omega_D$ for
$V_0=0.1$. (c) Same as (a) but for $V_0=0.35$ (d) Same as (b) but
for $V_0=0.35$. For all plots all energies and frequencies are
measured in units of ${\mathcal J}_0$, $\hbar$ is set to unity, and
the chain length is $L=14$. See text for details.} \label{fig2}
\end{figure}

Using Eqs.\ \ref{wavpert} and \ref{wavexact}, we find the
wavefunction overlap between the exact and perturbative
wavefunctions to be
\begin{eqnarray}
F[|\psi_0\rangle] &=& |_{\rm exact}\langle \psi(T)|\psi(T)
\rangle_{\rm
pert}| \nonumber\\
&=& \Big| \sum_{mn} d_m^{\ast}(T) c_n \Lambda_{mn}
e^{-i(\epsilon_n^F-\epsilon_m) T} \Big| \nonumber\\
C_{\rm av} &=& -\sum_{|\psi_0\rangle} \ln(1-F[|\psi_0\rangle])
\label{feq1}
\end{eqnarray}
where $\Lambda_{mn}= \langle \phi_m|\chi_n\rangle$ denotes the
overlap between the Floquet and the exact eigenstates and the sum
over $|\psi_0\rangle$ indicates sum over random initial states
chosen from the Hilbert space of $H$ (Eq.\ \ref{fermham1}). A plot
of $C_{\rm av}$ as a function of $\omega_D$ for $V_0/{\mathcal
J}_0=0.1$ is shown in Fig.\ \ref{fig2}(a); the corresponding plot
for $V_0/{\mathcal J}_0=0.35$ is shown in Fig.\ \ref{fig2}(c). Here
we have obtained $C_{\rm av}$ by averaging over $50$ random initial
states chosen from the Hilbert space of $H$ (Eq.\ \ref{fermham1})
with total occupation set to half filling $N=L/2$. We have checked
that $\sigma_C = \sum_{|\psi_0\rangle} (-\ln(1-F[|\psi_0\rangle]
-C_{\rm av})^2/C_{\rm av} \ll 1$ as expected from standard
typicality arguments \cite{typref1}. We have also computed analogous
quantity $C_{\rm av}^m$, where $U_{\rm pert}(T,0)$ in Eq.\
\ref{uevolpert1} is replaced by its counterpart from the Magnus
Floquet Hamiltonian $H_F^{\rm magnus}= H_F^{(1)\rm magnus}$. The
plot show that $C_{\rm av} \ge 3$ for $\omega_D \ge V_0=0.1$; the
corresponding quantity for Magnus displays a significantly lower
value for all $\omega_D/{\mathcal J}_0 \le 2$. Fig.\ \ref{fig2}(b)
and (d) shows similar plots $C_{\rm prod}$ obtained using a product
initial state (which shall be used as a starting state for studying
dynamical localization in this model in Sec.\ \ref{dynloc})
\begin{eqnarray}
|\psi_p\rangle &=& |n_1=1, .. n_{\ell}=1, n_{\ell+1}=0 ...
n_L=0\rangle, \label{prodin}
\end{eqnarray}
where $\ell=L/2$ for even $L$ and $\ell=(L-1)/2$ for odd $L$. We
find that $C_{\rm prod}$ also shows analogous behavior. Our results
thus indicate that $H_F$ obtained using FPT provides a much better
approximation than its counterpart obtained using Magnus expansion
to exact numerics for all $\omega_D/ V_0 \ge 1$ and for
$V_0/{\mathcal J}_0 \ll 1$.

\begin{figure}
\rotatebox{0}{\includegraphics*[width=0.49\linewidth]{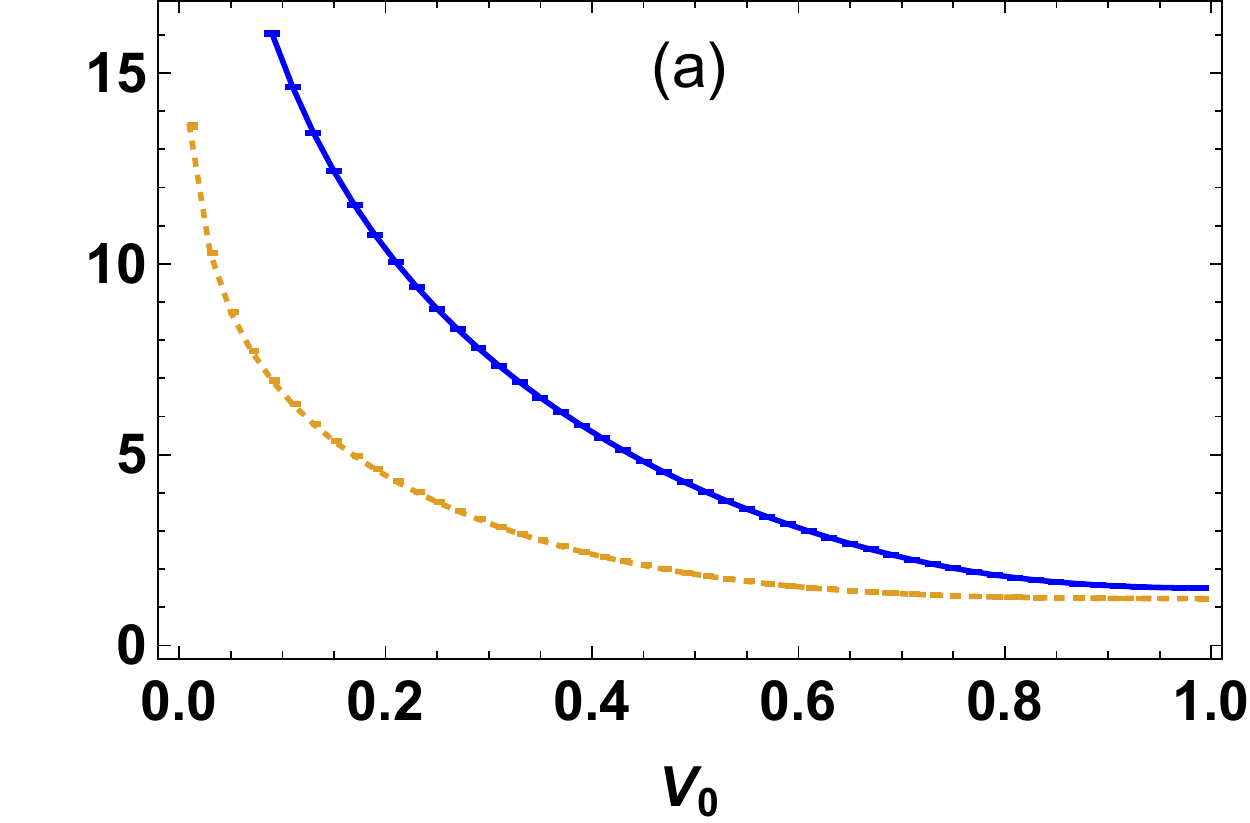}}
\rotatebox{0}{\includegraphics*[width=0.49\linewidth]{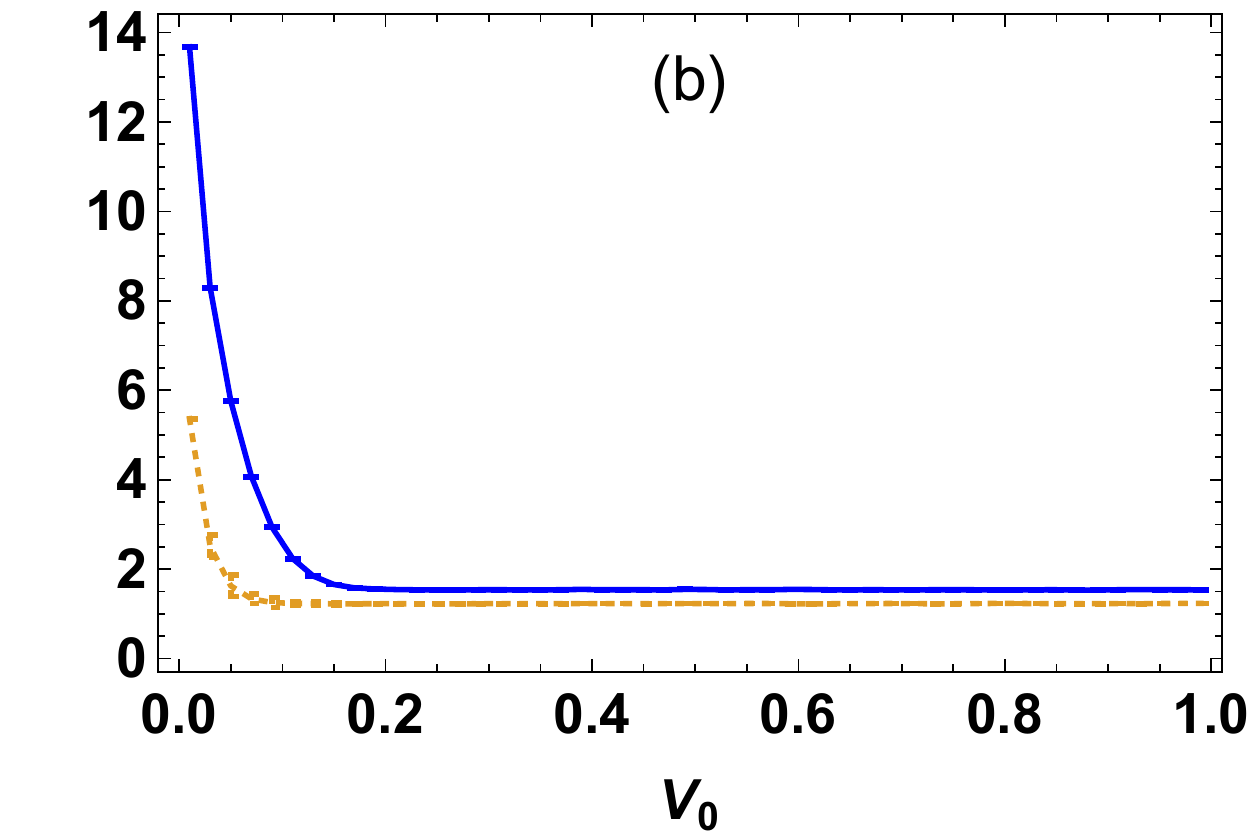}} \\
\rotatebox{0}{\includegraphics*[width=0.49\linewidth]{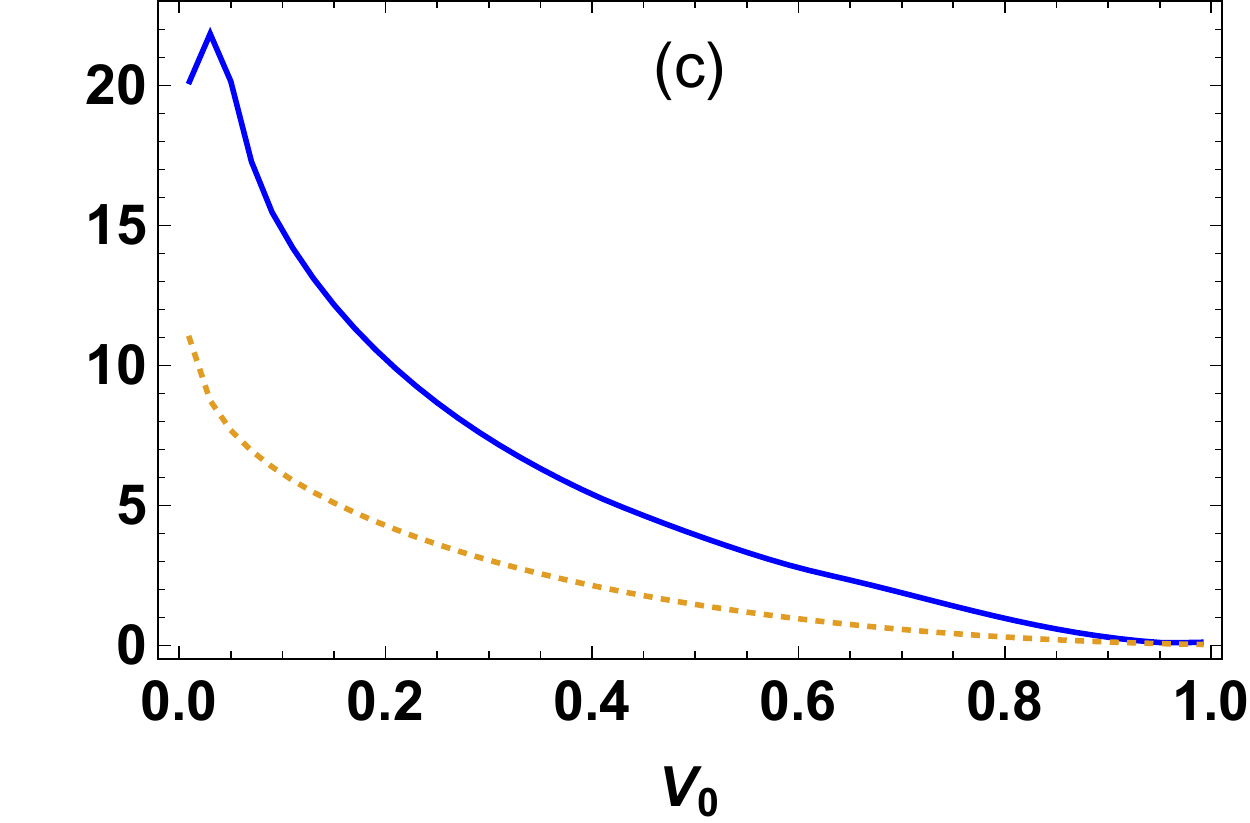}}
\rotatebox{0}{\includegraphics*[width=0.49\linewidth]{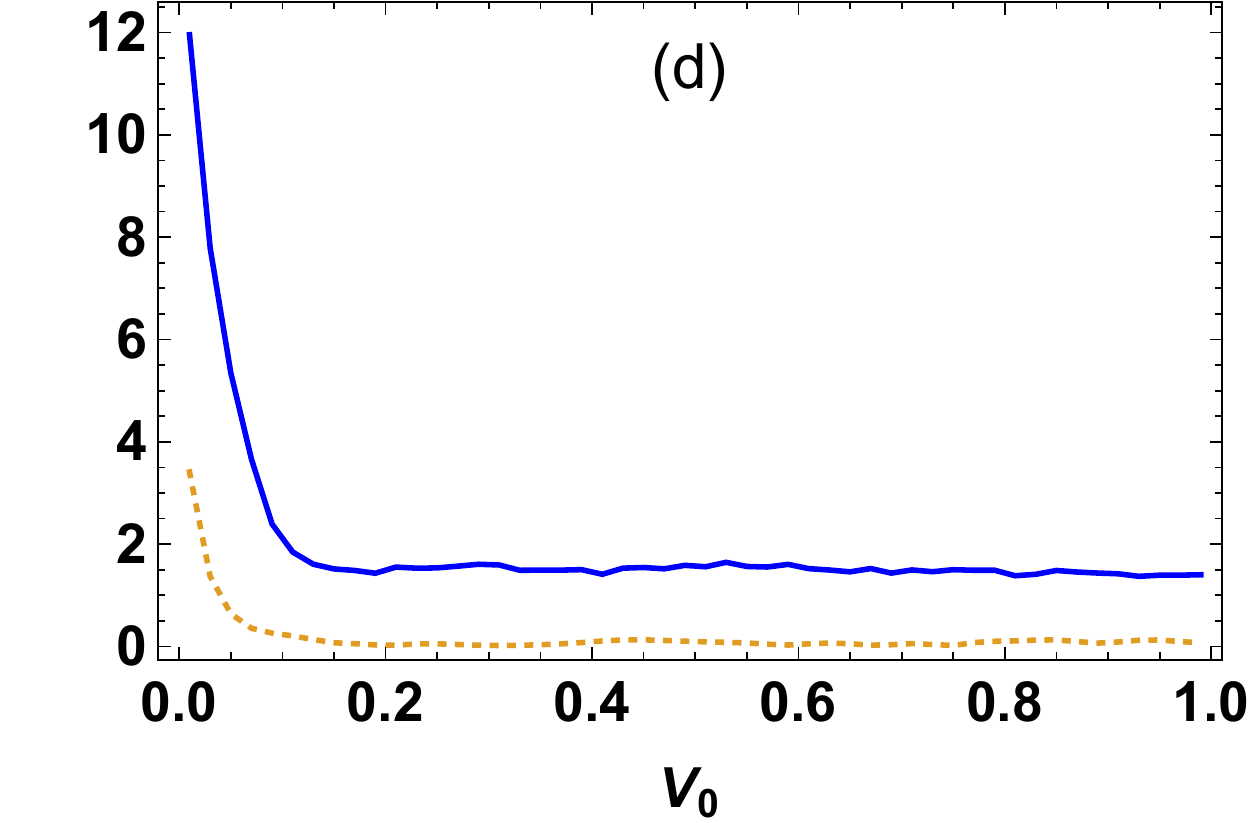}}
\caption{(a) Plot of $C_{\rm av}$ (blue solid line) and $C_{\rm
av}^m$ (yellow dotted line) for $\omega_D=1$ as a function of $V_0$.
(b) Same as (a) but for $\omega_D=0.1$ (c) Plot of $C_{\rm prod}$
(blue solid line) and $C_{\rm prod}^m$ (yellow dotted line) for
$\omega_D=1$ as a function of $V_0$. (d) Same as (c) $\omega_D=0.1$.
For all plots all energies and frequencies are measured in units of
${\mathcal J}_0$, $\hbar$ is set to unity, and the chain length is
$L=14$. See text for details.} \label{fig3}
\end{figure}

Fig.\ \ref{fig2} also brings out the perturbative nature of our
results; we find, by comparing Fig.\ \ref{fig2}(a) and (b) with
Fig.\ \ref{fig2} (c) and (d) respectively, that both $C_{\rm av}$
and the fidelity for the product state shows larger value for
$V_0/{\mathcal J}=0.1$ for same $V_0/\omega_D$. To elucidate this
point further, we plot $C_{\rm av}$ as a function of $V_0/{\mathcal
J}_0$ in Fig.\ \ref{fig3}(a) and (b) for $\omega_D/J_0=1$ and $0.1$
respectively. Analogous plots for the product state is shown in
Fig.\ \ref{fig2}(c) and (d). From these plots we find that both
$C_{\rm av}$ and $C_{\rm prod}$ decreases with increasing
$V_0/{\mathcal J}_0$ and that such a decrease is more rapid at lower
frequencies. This points out that our method provide a much more
accurate description compared to the Magnus expansion for high and
intermediate frequencies and low interaction strength; however, it
fails for large interaction strength and low frequencies, as is
expected within our perturbative approach.

\section{Application to dynamics}
\label{secapp}

In this section, we shall discuss several applications of the FPT
developed earlier. In Sec.\ \ref{ssapp}, we discuss the approach of
the driven interacting fermionic chain to its steady state while in
Sec.\ \ref{dynloc}, we discuss transport in such driven system with
emphasis on the phenomenon of dynamical localization.

\subsection{Approach to the steady state}
\label{ssapp}

The approach to the steady state of a driven periodic system can be
studied from its Floquet Hamiltonian. To this end, we follow Ref.\
\onlinecite{rigol1} and consider a quantity $Q$ defined as
\begin{eqnarray}
Q = \frac{ \langle \psi(n_0 \to \infty)|H_{\rm av}|\psi(n_0 \to
\infty) \rangle - \langle H_{\rm av} \rangle_{\beta \to 0}}{\langle
H_{\rm av}\rangle_{\beta \to 0}- \langle \psi(t=0)|H_{\rm
av}|\psi(t=0)\rangle } \label{qdef1}
\end{eqnarray}
Here $H_{\rm av} = \int_0^T H(t) dt/T = H_1$ is the average
Hamiltonian, $\beta=(k_B T_0)^{-1}$ is the inverse temperature,
$k_B$ is the Boltzmann constant, $|\psi( n_0\to \infty)\rangle$
indicates the steady state wavefunction, and $\langle H_{\rm
av}\rangle_{\beta \to 0}$ and $\langle \psi(t=0)|H_{\rm
av}|\psi(t=0)\rangle$ denotes the values of $H_{\rm av}$ in the
infinite temperature and the initial states respectively. We note
that $Q=0$ if the steady state reaches the infinite temperature
value; in contrast $Q \simeq-1$ if the system does not respond to
the drive and stays close to its initial state. Thus for all
starting states $-1 \le Q \le 0$; its intermediate values signifies
finite-temperature steady states as pointed out in Ref.\
\onlinecite{rigol1}. Eq.\ \ref{qdef1} holds for pure
initial states; its counterpart for mixed states represented by a
density matrix $\rho$ can be easily obtained by the substitution
$\langle \psi|H_{\rm av}|\psi\rangle \to {\rm Tr}[ \rho H_{\rm
av}]$.

To compute $Q$ using the Floquet Hamiltonian derived from FPT and
for a pure initial state, we note that in terms of the Floquet
eigenvalues $\epsilon_m^F$ and eigenfunctions $|\chi_m\rangle$, the
wavefunction after $n_0$ drive cycles can be written as
$|\psi(n_0T)\rangle = \sum_m c_m \exp[-i n_0 \epsilon_m^F T]
|\chi_m\rangle$ where $c_m$ denotes the overlap between the initial
and the $m^{\rm th}$ Floquet eigenstate. Using this, we find
\begin{eqnarray} \langle \psi(n_0 T)\rangle|H_{\rm av}| \psi(n_0
T)\rangle &=& \sum_{m_1,m_2} c_{m_1}^{\ast} c_{m_2} e^{i n_0 T (
\epsilon_{m_1}^F-
\epsilon_{m_2}^F)} \nonumber\\
&& \times \langle \chi_{m_1}|H_{av}| \chi_{m_2}\rangle \label{havexp1}
\end{eqnarray}
In the steady state, the contribution to the sum comes from diagonal
matrix elements and those off-diagonal elements for which the states
$|\chi_{m_1}\rangle$ and $|\chi_{m_2}\rangle$ are degenerate. Thus
one finds
\begin{eqnarray} \langle \psi(\infty)\rangle|H_{\rm av}|
\psi(\infty)\rangle &=& \sum_{m_1}
|c_m|^2 \langle \chi_{m}|H_{av}| \chi_{m}\rangle  \label{havexp2} \\
&& + \sum'_{m_1,m_2} c_{m_1}^{\ast} c_{m_2} \langle \chi_{m_1}|H_{av}|
\chi_{m_2}\rangle\nonumber
\end{eqnarray}
where $\sum'$ denotes sum over degenerate states. The computation of
this quantity using ED involves finding the wavefunction after $n_0$
drive cycles and computing expectation of $H_1$ using this
wavefunction. The steady state value of this quantity yields
$\langle \psi(\infty)\rangle|H_{\rm av}| \psi(\infty)\rangle_{\rm
exact}$.

 In contrast for a mixed thermal initial state, one
needs to invoke its density matrix $\rho_{\rm init}=
|\psi(0)\rangle\langle \psi(0)| =\sum_m \exp[-\beta \epsilon_m^1]
|\zeta_m\rangle\langle \zeta_m|/Z$, where $Z=\sum_{m} \exp[-\beta
\epsilon_m^1]$ is the partition function, $\epsilon_m^1$ and
$|\zeta_m\rangle$ denotes the $m^{\rm th}$ eigenvalue and
eigenvector of $H_{\rm av}=H_1$ respectively, $\beta =1/(k_B T_0)$
is the inverse temperature, and $k_B$ is the Boltzmann constant.
Using $\rho(n_0 T) = U(n_0 T,0) \rho(0) U^{\dagger}(n_0 T,0)$, we
find after a straightforward calculation
\begin{eqnarray}
\langle H_{\rm av} \rangle_{n_0 \to \infty} &=& \sum_{k,m,p} |b_{k
m}|^2 \frac {e^{-\beta \epsilon_m^1}}{Z} |b_{k p}|^2 (H_{\rm
av})_{pp} \label{deq}
\end{eqnarray}
where $b_{mn} = \langle \chi_m |\zeta_n\rangle$. For such states,
exact numerics using ED requires solution of equation of motion for
matrix elements of the density matrix of the system and is
computationally intensive.

The computation of expectation values of $H_1$ in the infinite
temperature and initial state, involves obtaining $\epsilon_m^1$ and
$|\zeta_m\rangle$ by numerically diagonalizing $H_1$ using ED. One
can then use this basis to obtain these quantities as
\begin{eqnarray}
\langle H_{\rm av} \rangle_{\beta=0} &=& \frac{1}{\mathcal{D}}\sum_m
\epsilon_{m}^1 \label{havexp3} \\
{\rm Tr}[\rho_{\rm init} H_{\rm av}]&=& \sum_m  e^{-\beta
\epsilon_{m}^1} \epsilon_{m}^1/Z \nonumber
\end{eqnarray}
where we have taken mixed initial state with temperature $T_0$ and $\mathcal{D}$ is the Hilbert space dimension. An
analogous expression for $\langle \psi(0)|H_{\rm av}|\psi(0)\rangle$
starting from the product initial state can also be easily obtained
and is given by $\langle \psi(0)|H_{\rm av}|\psi(0)\rangle= \sum_m
g_m \epsilon_{m}^1$ where $g_m=\langle \zeta_m|\psi_p\rangle$.
Substituting these results in Eq.\ \ref{qdef1}, one can numerically
obtain $Q$ using FPT for both thermal mixed and pure initial
states.

\begin{figure}
\rotatebox{0}{\includegraphics*[width=0.49\linewidth]{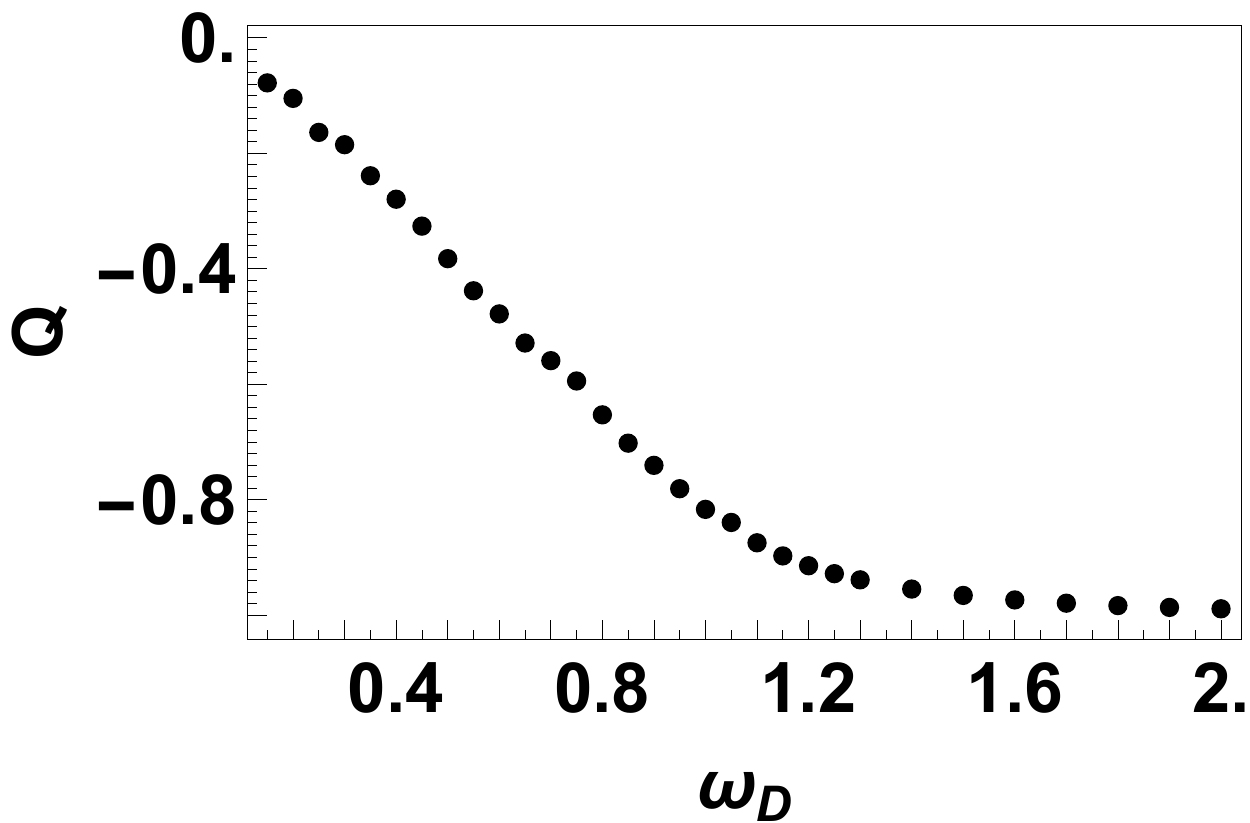}}
\rotatebox{0}{\includegraphics*[width=0.49\linewidth]{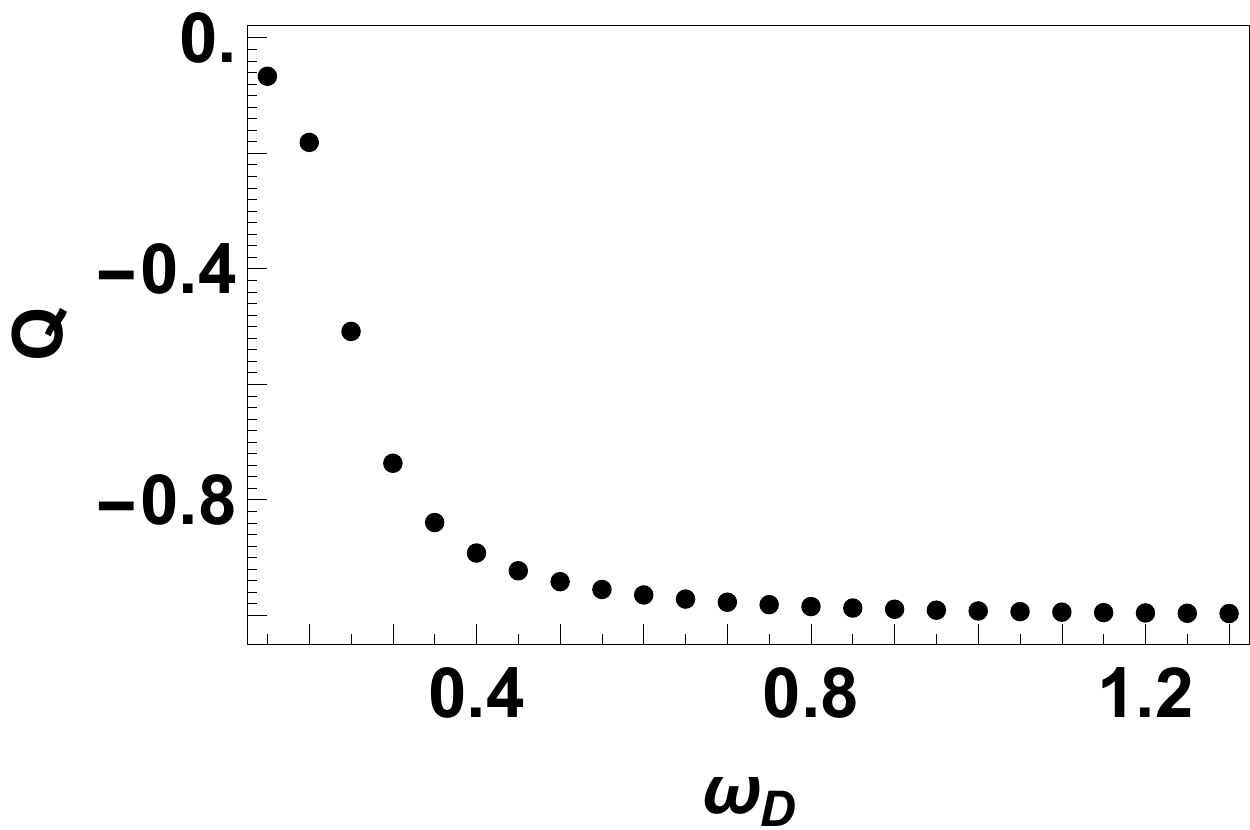}}
\caption{(a) Plot of $Q$ as a function of $\omega_D$ for $V_0=0.1$
showing approach to the infinite temperature steady state. The left
panel corresponds to a thermal initial state with $k_B T_0=0.01$
while the right panel corresponds to the initial state
$|\psi_p\rangle$ (Eq.\ \ref{prodin}). For all plots all energies and
frequencies are measured in units of ${\mathcal J}_0$, $\hbar$ is
set to unity, and the chain length is $L=16$. See text for details.}
\label{fig4}
\end{figure}

The results of such computation for finite chain $L=16$ are shown in
Fig.\ \ref{fig4}. The left panel of Fig.\ \ref{fig4} shows $Q$ as a
function of $\omega_D$ starting from a low temperature ($k_B
T_0=0.01{\mathcal J}_0$) thermal density matrix while the right
panel corresponds to the initial product state given by Eq.\
\ref{prodin}. For both cases, we find that $Q \simeq -1$ at high
frequency showing that the system does not absorb energy in the high
frequency regime. This is consistent with the fact that in this
regime $H_F \simeq H_1= H_{\rm av}$ so that $[U, H_{\rm av}] \simeq
0$. In contrast, in the low frequency regime $\omega_D \ll V_0$, the
system reaches in the infinite temperature steady state and $Q \to
0$. In between, for a wide range of frequency $V_0 \le \hbar
\omega_D \le {\mathcal J}_0$, the system reaches subthermal (for the
initial thermal density matrix) or superthermal (for the initial
product state) steady states (for finite-size chain) with $-1\le Q
\le 0$.

\begin{figure}
\rotatebox{0}{\includegraphics*[width=0.49 \linewidth]{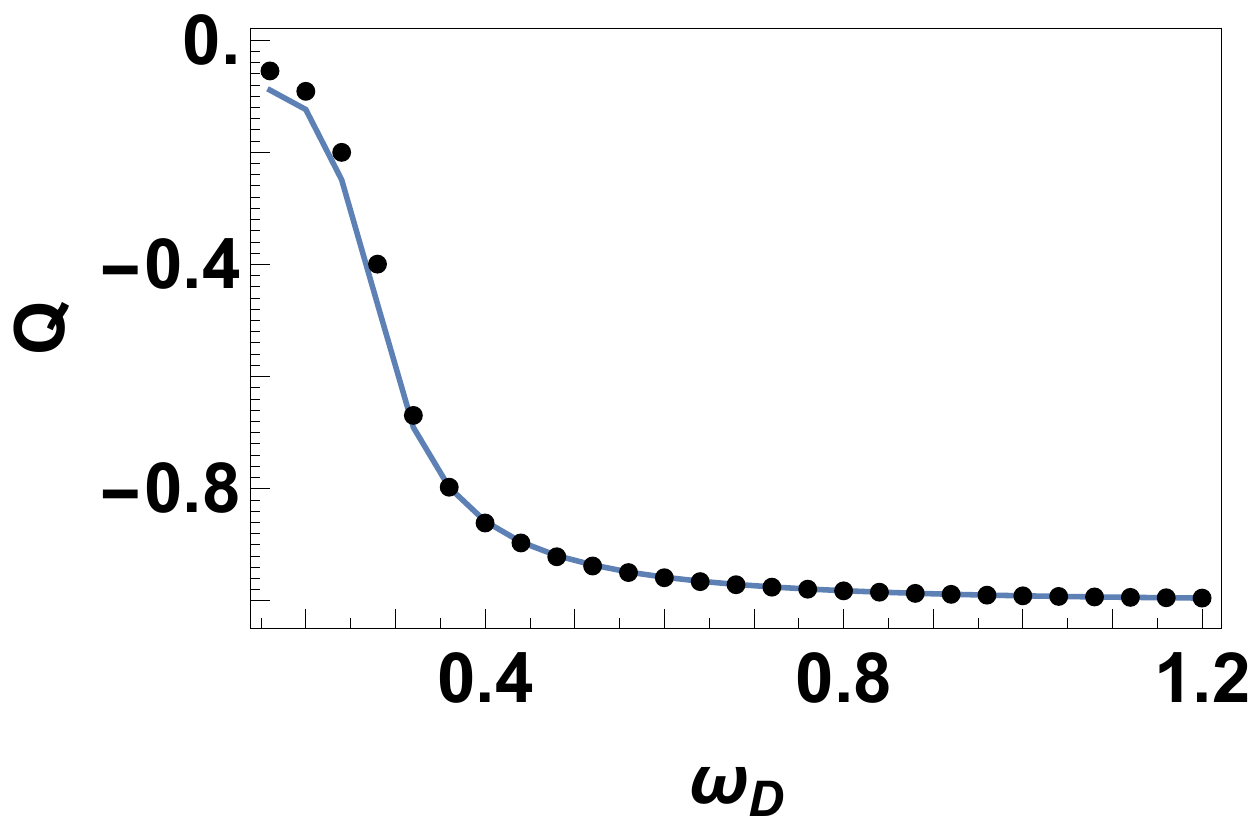}}
\rotatebox{0}{\includegraphics*[width=0.49 \linewidth]{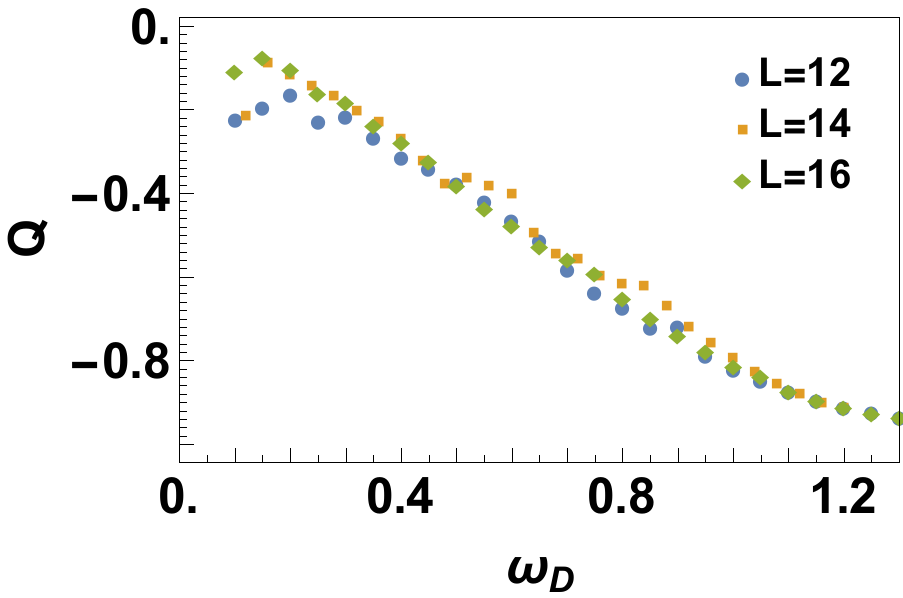}}
\caption{Left Panel: Plot of $Q$ as a function of $\omega_D$
starting from $|\psi_p\rangle$ for $L=14$ and $V_0=0.1$. The black
dots correspond to FPT results while the blue line indicates exact
numerics using ED. Right panel: Plot of $Q$ as a function of
$\omega_D$ starting from the thermal mixed state ($k_B T_0=0.01$)
for $V_0=0.1$ and different system sizes as indicated. All energies
and frequencies are measured in units of ${\mathcal J}_0$ and $\hbar$
is set to unity. See text for
details.} \label{fig5}
\end{figure}

To verify the accuracy of FPT, we compute $Q$ using exact numerics
and compare it with its counterpart obtained using FPT for $L=14$
and starting from $|\psi_p\rangle$. The result shown in the left
panel of Fig.\ \ref{fig5} indicates that FPT provides accurate
description of the behavior of $Q$ for all frequencies $\omega_D \ge
V_0$. This property is contrasted with $Q$ obtained from Magnus
expansion; since $H_F=H_1$, $Q=-1$ for all $\omega_D$ in this case
and the crossover can never be captured. The right panel of Fig.\
\ref{fig5} shows the system size dependence of $Q$ as obtained using
FPT for $L=12,\, 14\, {\rm and}\,16$ starting from the thermal
initial state with $k_B T_0=0.01 {\mathcal J}_0$. We find that the
broad crossover region at intermediate frequencies is almost
independent of system size in this case. This may indicate that such
a phenomenon will be observed as prethermal behavior for
thermodynamic chains; we shall discuss this issue in details in the
next section.


Finally, we compute the Shannon entropy corresponding to $U$. To
this end, we numerically compute the overlap $c_{n}^m = \langle
\zeta_m|\chi_n\rangle$ between the eigenstates $|\zeta_m\rangle$ of
$H_{\rm av}= H_1$ computed using ED and $|\chi_n\rangle$ of $H_F$
obtained using second order FPT. In terms of the Shannon entropy $S$
is given by
\begin{eqnarray}
S &=& \sum_n S_n /S_0,\,\, S_n= -\sum_m |c_n^m|^2 \ln |c_n^m|^2
\label{sentr1}
\end{eqnarray}
where $S_0 = \ln 0.48 {\mathcal D}$ is the ETH predicted
infinite-temperature steady state value of $S$ for a circular
orthogonal ensemble (COE) and ${\mathcal D}$ is the Hilbert space
dimension\cite{rigol1}.

\begin{figure}
\rotatebox{0}{\includegraphics*[width=0.49\linewidth]{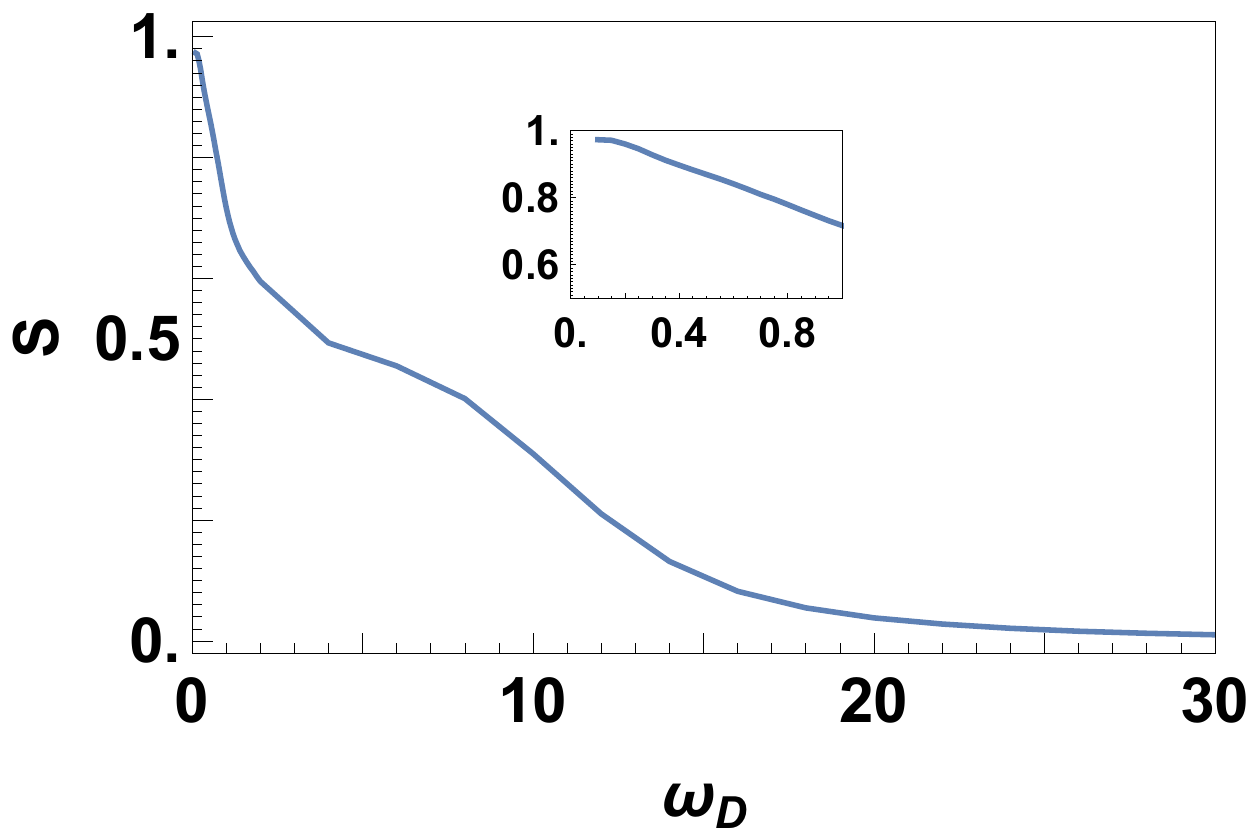}}
\rotatebox{0}{\includegraphics*[width=0.49\linewidth]{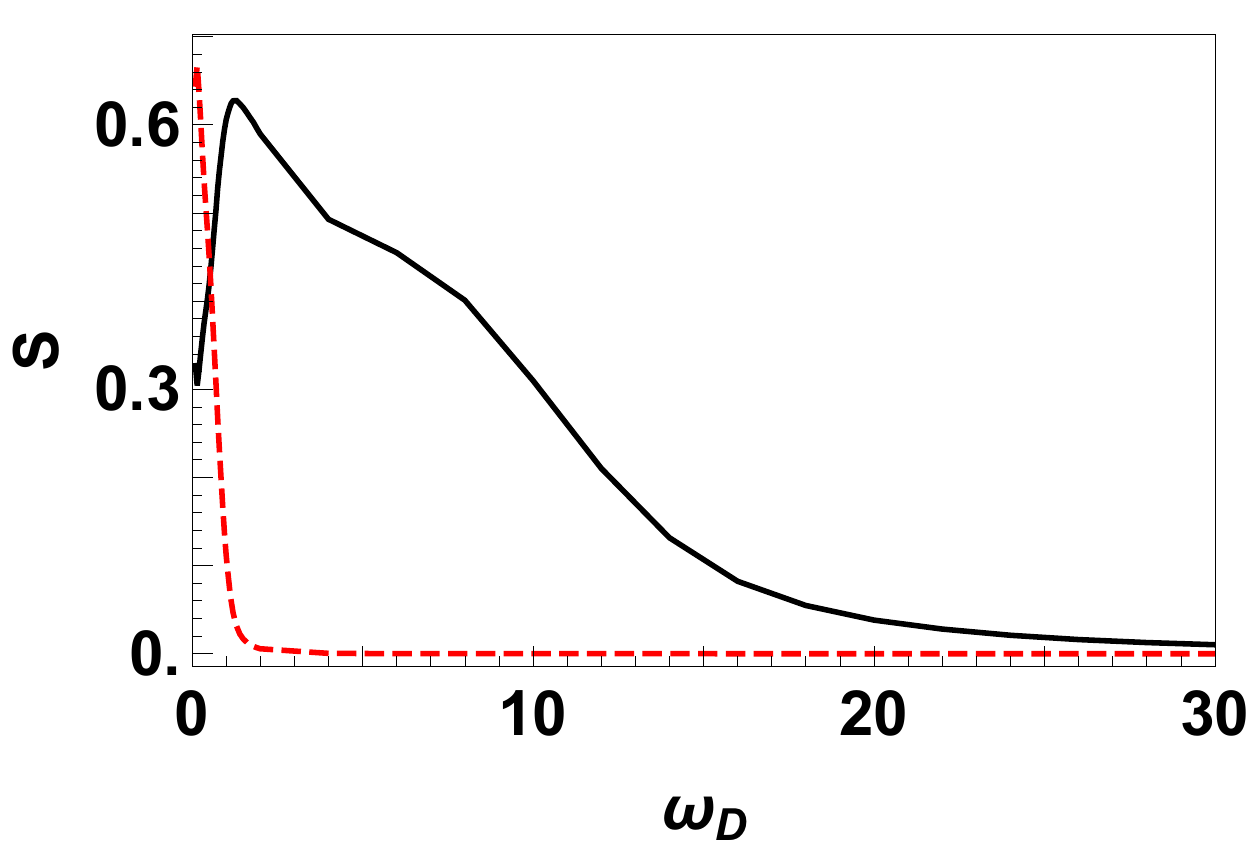}}
\caption{Left Panel: Plot of $S$ as a function of $\omega_D$ for
$V_0=0.1$. Right panel: Plot of interband (red dotted line) and
intraband (black solid line) contribution to $S$ as a function of
$\omega_D$. All energies and frequencies are measured in units of
${\mathcal J}_0$, $\hbar$ is set to unity, and the chain length is
$L=16$. See text for details.} \label{fig6}
\end{figure}

A plot of $S$ as a function of the drive frequency $\omega_D$ is
shown in Fig.\ \ref{fig6}. We find that $0 \le S \le 1$ for our
system. At large drive frequency $S \to 0$ since $H_F \simeq H_{\rm
av}=H_1$ in this limit. $S$ increases towards its COE predicted
value as the drive frequency is reduced and attains this value
around $\hbar \omega_D \simeq 2V_0$ as seen from the inset of left
panel of Fig.\ \ref{fig6}. This increase occurs with two distinct
slopes. At higher frequencies, $S$ increases with a lower slope;
this changes to a sharper rise for $\hbar \omega_D/{\mathcal J}_0
\le 1$. To explain this feature, we show, in the right panel of
Fig.\ \ref{fig6}, the contribution of inter- and intra-band overlaps
to $S$. We find that high $\hbar \omega_D \ge {\mathcal J}_0$, the
entire contribution to $S$ comes from the intra-band overlaps
$c_{n}^m$ with $n$ and $m$ being states in the same nearly flat
bands; $c_{n}^m$ between states where $n$ and $m$ belongs to
different flat bands vanishes in this region. As the frequency
decreases the eigenstates of $H_F$ starts to delocalize and around
$\hbar \omega_D \simeq {\mathcal J}_0$, they have overlap with
multiple flat-band eigenstates of $H_1$. This leads to additional
contribution to $S$ and leads to its sudden sharp increase as can be
seen from right panel of Fig.\ \ref{fig6}. We note that the presence
of such multiple slope of $S$ as a function of $\omega_D$ is a
consequence of flat band structure of $H_1$.

We find that for $V_0 \le \omega_D$, where we can trust the
prediction of FPT, $S \le 1$ for a wide range of drive frequencies;
this further confirms the presence of subthermal or superthermal
steady states in these driven finite sized fermionic chains. We note
here that computation of $S$ necessitates inputs from FPT; for the
continuous drive protocol that we study here, it is quite difficult
to compute eigenvectors of $U$ reliably using ED via trotterization
of ${\mathcal T}\exp[-i \int_0^T dt H(t)]$. Thus one can not easily
compute $S$ exactly in contrast to the case of pulsed protocols as
done in Ref.\ \onlinecite{rigol1}. Finally, we note the Magnus
expansion for which $H_F=H_1$ at all $\omega_D$ predicts $S=0$ at
all drive frequencies.

\subsection{Dynamical localization}
\label{dynloc}

In the absence of interaction, the driven fermionic chain described
by $H_0(t)$ (Eq.\ \ref{fermham1}) exhibits exact dynamical
localization at stroboscopic times. This is easily seen by noting
$U_0(T,0)=1$ (Eq.\ \ref{evolzero}) so that $|\psi(n_0 T)\rangle =
|\psi(0)\rangle$ for all $T$ and $n_0$. At intermediate times, an
initial state evolves; however it exhibits localization. To see
this, let us consider the initial state $|\psi_p\rangle$ (Eq.\
\ref{prodin}). For $V_0=0$ and $d=1$, one can obtain an exact
expression for the fermionic annihilation operator
\cite{antal1,antal2,eisler1}
\begin{eqnarray}
c_{k}(t) &=& U_{k}^{\dagger}(t,0) c_{k}(0) U_{k}(t,0) \nonumber\\
&=& e^{-i {\mathcal J}_0 \sin(\omega_D t) \cos k/(\hbar \omega_D)}
c_k(0)\label{opeq2}
\end{eqnarray}
In real space, one can thus write
\begin{eqnarray}
c_j(t) = \sum_{j'} J_{j-j'}(\Lambda(t)) i^{j-j'} c_{j'}(0)
\label{opeq3}
\end{eqnarray}
where $j$ and $j'$ are site indices and $\Lambda(t)= {\mathcal J}_0
\sin(\omega_D t)/(\hbar \omega_D)$. The fermionic density for the
state $|\psi_p\rangle$ at any time $t$ for $j>0$ is thus given by
\begin{eqnarray}
n_j(t) &=& \sum_{j'>j} J^2_{j'}(\Lambda(t)). \label{fermden1}
\end{eqnarray}
We now ask the question: at what time, within a single drive cycle
($t \le T$) do the fermions reach a specific site $j_0$. An analytic
estimate of this time could be obtained by noting that $J_j(x)$
remains close to zero for $x \le j$; it becomes finite when $x\ge
j$. Thus we find that the time $t_0$ taken by the fermions to reach
a distance $j_0 = j-L/2$ to the right of the density front centered
at $j=L/2$ can be estimated to be (the lattice spacing is set to
unity) $\Lambda(t_0) \simeq j_0$. This immediately tells us that for
any protocol for which $\Lambda(t)$ is a bounded function of time,
there may not exist any real-valued solution of $t_0$ for large
enough $j_0$. Thus the fermions may never reach a site sufficiently
far away from the edge of the density front at $j=L/2$. Indeed, for
the sinusoidal protocol we use, one has
\begin{eqnarray}
t_0 &=& \omega_D^{-1} \arcsin(j_0 \hbar \omega_D/{\mathcal J}_0)
\label{timeest2}
\end{eqnarray}
Eq.\ \ref{timeest2} has no real solution for $t_0$ for $j_0 >
{\mathcal J}_0/(\hbar \omega_D)$ which indicates that fermions will
never reach a site $j_0 > {\rm Int}[{\mathcal J}_0/(\hbar
\omega_D)]$, where ${\rm Int}[x]$ denotes the nearest integer to
$x$. Also,this indicates that a driven non-interacting chain will
exhibit perfect dynamic localization at all times for $\hbar
\omega_D > {\mathcal J}_0$.

The presence of interaction is expected to delocalize the fermion.
To investigate this effect, we now consider the steady behavior of
two correlation functions \cite{dl2} 
\begin{eqnarray}
N_{\rm av}(T) &=&  \frac{4}{L}   \sum_j \langle (n_j-1/2)\rangle ^2 \label{opdef1} \\
M(T) &=& 1- \frac{1}{{\mathcal L}_0} \sum_j
j_d^2\langle(n_j-1/2)\rangle^2  \nonumber
\end{eqnarray}
where $j_d= j-L/2[(L-1)/2]$ for even[odd] $L$, ${\mathcal L}_0 =
\sum_{j=1,l} j_d^2/4$ is the normalization, and the average is taken
with respect to the steady state reached when the system is driven
with frequency $\omega_D$ and $|\psi_p\rangle$ is chosen to be the
initial state. In terms of the Floquet eigenvectors, one can write
\begin{eqnarray}
&&N_{\rm av}(T) =\sum_j  \frac{4}{L}( \sum_n |c_n|^2 \langle
\chi_n|(n_j-1/2)|\chi_n\rangle)^2 \label{opdef2}\\
&& M(T) =1- \frac{1}{{\mathcal L}_0}\sum_j j_d^2( \sum_n [ |c_n|^2
\langle \chi_n|(n_j-1/2)|\chi_n\rangle ])^2  \nonumber
\end{eqnarray}
where $c_n=\langle \chi_n|\psi_p\rangle$. We note that for the
initial state, $(4/L) \sum_j (\langle
\psi_p|(n_j-1/2)|\psi_p\rangle)^2=1$ while for the uniform state it
vanishes. Thus the deviation of $N_{\rm av}(T)$ from unity denotes
delocalization. In addition, we have used the fact that for
$|\psi_p\rangle$, $\sum_j j_d^2(\langle \psi_p | (n_j-1/2)
|\psi_p\rangle)^2 = {\mathcal L}_0$. Thus $M(T) \to 0$ if the steady
state is close to the initial state by construction; its finite
value constitutes a signature of delocalization.

\begin{figure}
\rotatebox{0}{\includegraphics*[width=0.49\linewidth]{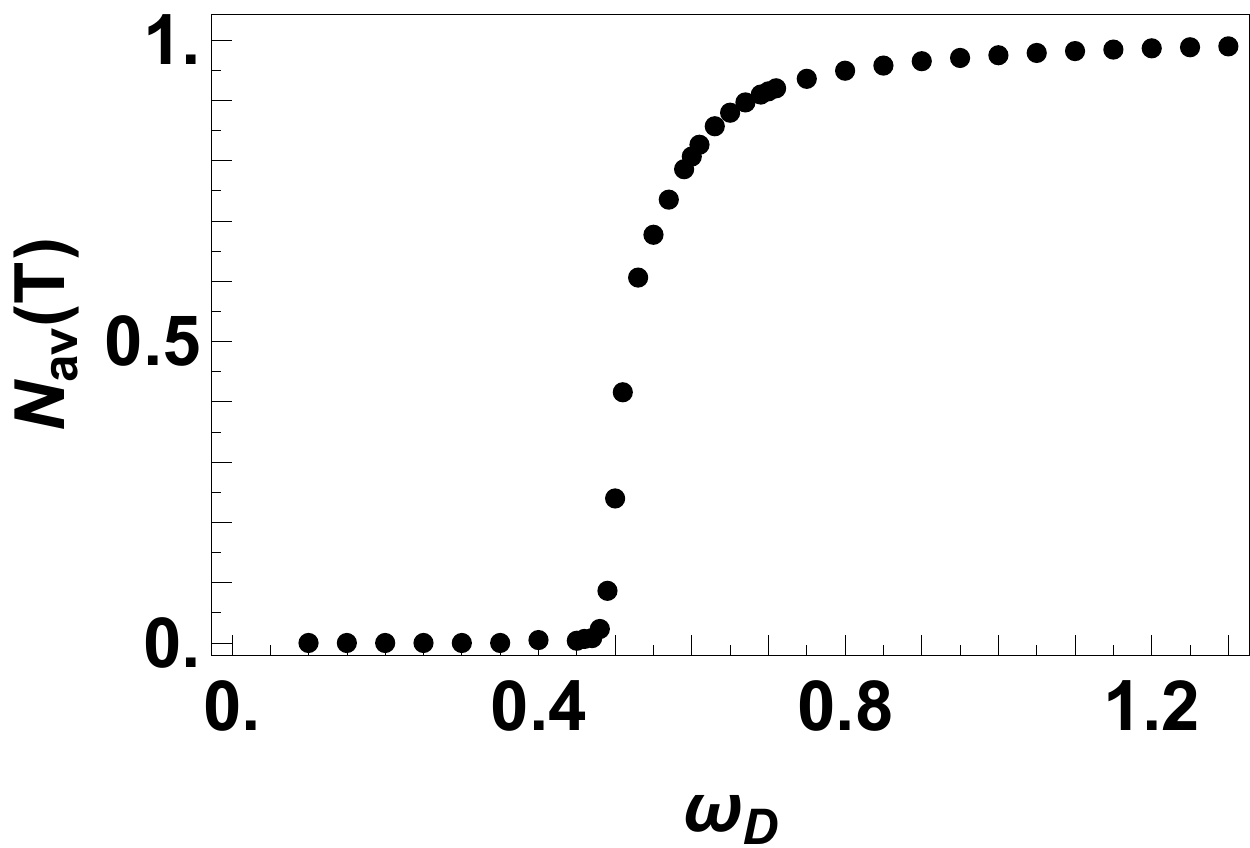}}
\rotatebox{0}{\includegraphics*[width=0.49\linewidth]{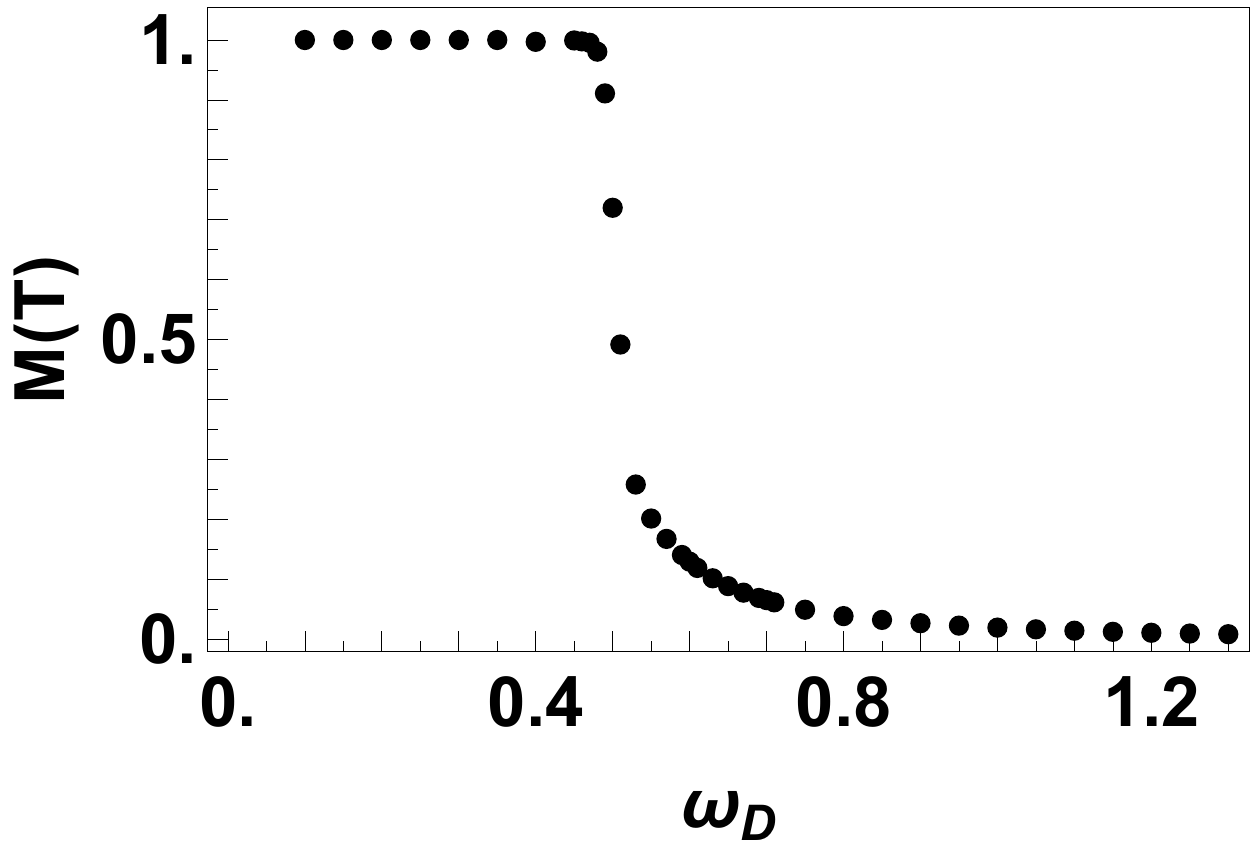}}
\rotatebox{0}{\includegraphics*[width=0.49\linewidth]{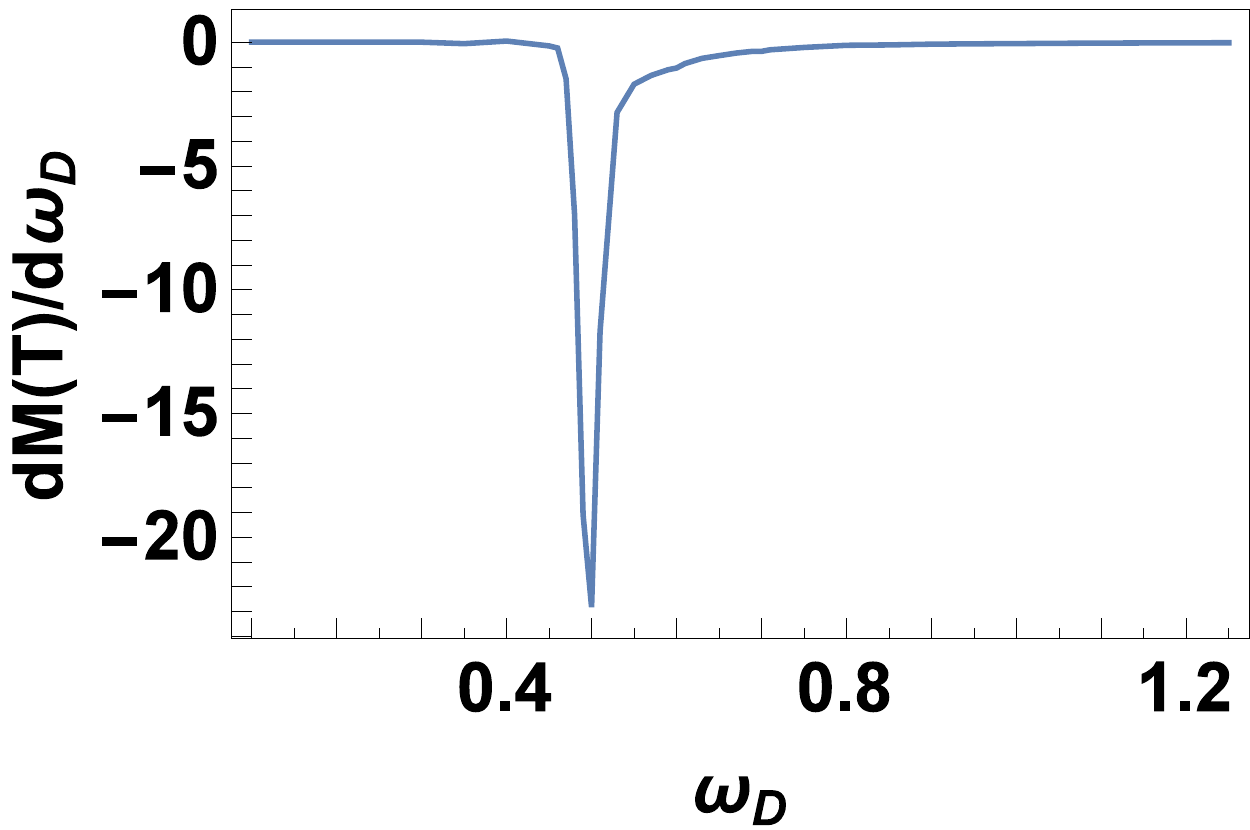}}
\rotatebox{0}{\includegraphics*[width=0.49\linewidth]{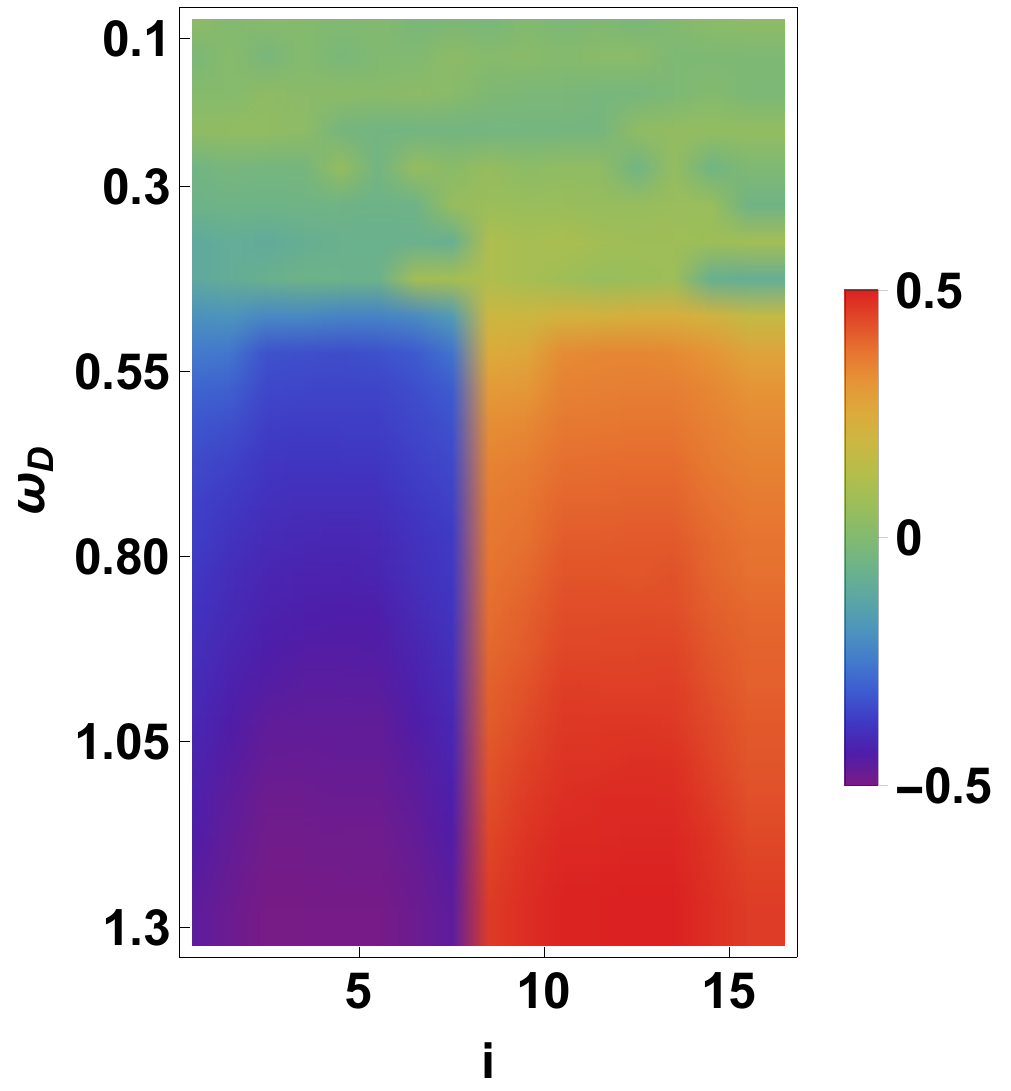}}
\caption{Top Left Panel: Plot of $N_{\rm av}(T)$ as a function of
$\omega_D$ for $V_0=0.1$. Top Right panel: Plot of $M(T)$ as a
function of $\omega_D$. Bottom left panel: Plot of
$dM(\omega_D)/d\omega_D$ as a function of $\omega_D$ quantifying the
rate of change in transport characteristics ($M$) of the steady
state. Bottom right panel: Plot of $\langle n_i(T) \rangle-0.5$ as a
function of the site index $i$ and frequency $\omega_D$. All plots
in the top panel show a clear crossover from delocalized to
localized regime around $\hbar \omega_D \simeq {\mathcal J}_0$. All
energies and frequencies are measured in units of ${\mathcal J}_0$,
$\hbar$ is set to unity, and the chain length is $L=16$. See text
for details.} \label{fig7}
\end{figure}

A plot of these quantities, using eigenfunctions obtained from
semi-analytic perturbative form of the Floquet Hamiltonian is shown
in the top panels of Fig.\ \ref{fig7} for $L=16$. We find that both
$N_{\rm av}$ and $M$ (Eqs.\ \ref{opdef1}) indicate a clear crossover
from localized to the delocalized steady states around
$\omega_D/{\mathcal J}_0 \simeq 1/2$. The bottom left panel shows a
plot of $dM/d\omega_D$ as a function of $\omega_D$ which brings out
the position of this crossover accurately. The bottom right panel of
Fig.\ \ref{fig7} shows the real-space density profile of the steady
state as a function of $\omega_D$ starting from $|\psi_p\rangle$. At
high-drive frequencies, one finds the steady state to have almost
the same density profile as the initial state; in contrast for
$\omega \simeq V_0$, the system is completely delocalized by the
time it reaches the steady state. In between there is a crossover
between the two states. We note that this crossover phenomenon can
also be understood from studying the structure of the Floquet
eigenstates. For $\omega_D \gg {\mathcal J}_0$, $H_F \simeq H_1$ so
that $[H_F, \hat n_j] \simeq 0$. Thus the density distribution does
not evolve significantly and the steady state remains close to the
initial state. However, for $\omega_D \le {\mathcal J}_0$, the
structure of $H_F^{(1)}$ changes; moreover, $H_F^{(2)}$ becomes
important. Thus in this regime $H_F$ does not commute with $n_j$ and
the system evolves to a steady state sufficiently different from the
initial state. In between a crossover between these two regimes
occur around $\omega_D \sim {\mathcal J}_0/2$ where the system
crosses over from localized to delocalized state for finite chains.
We note that one expects the steady state to be ETH predicted
thermal delocalized state for thermodynamic chains; thus such a
crossover is not expected in their steady states. However, as
discussed in Sec.\ \ref{diss}, the remnant of this behavior may be
seen as prethermal characteristics of such driven chains.

\begin{figure}
\rotatebox{0}{\includegraphics*[width=0.49\linewidth]{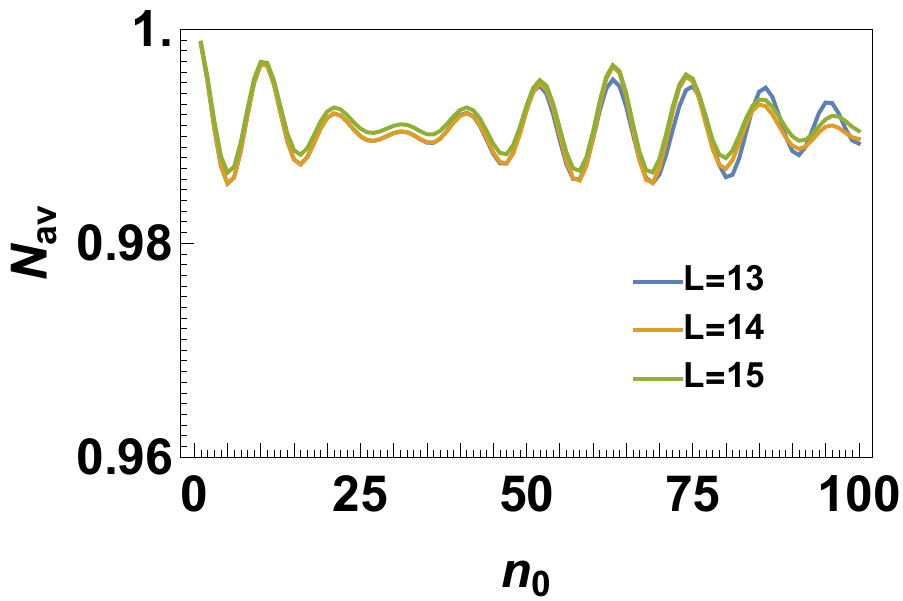}}
\rotatebox{0}{\includegraphics*[width=0.49\linewidth]{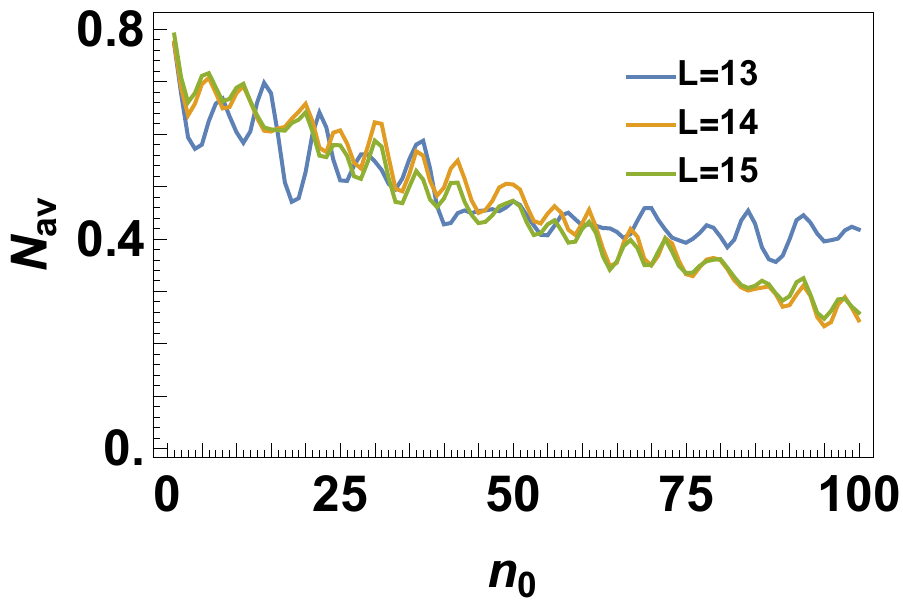}}
\rotatebox{0}{\includegraphics*[width=0.49\linewidth]{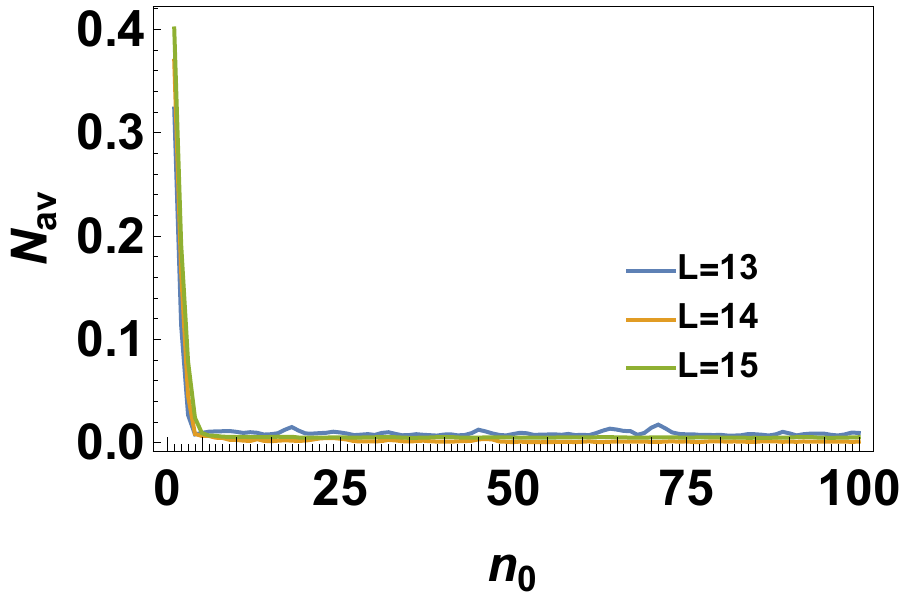}}
\rotatebox{0}{\includegraphics*[width=0.49\linewidth]{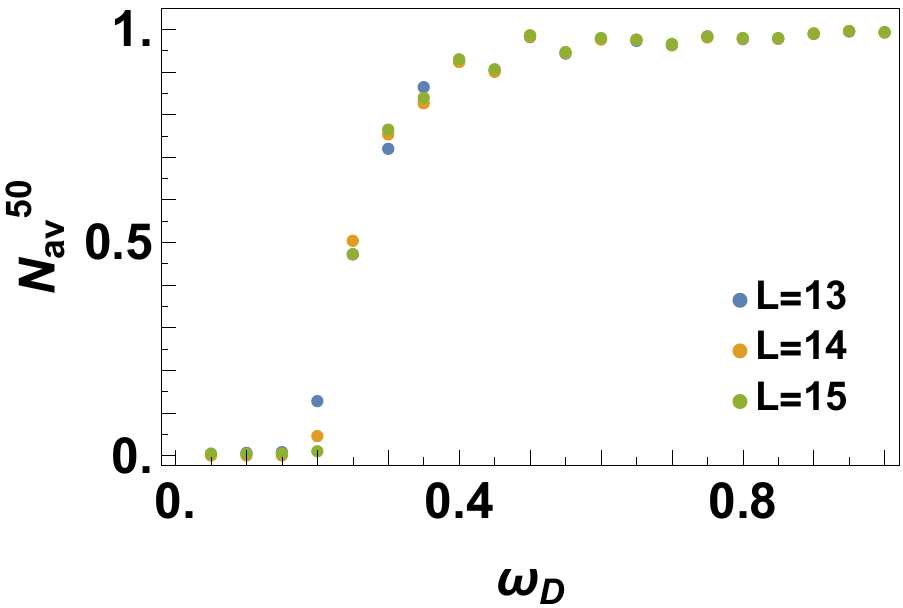}}
\caption{Plots of $N_{av}(n_0 T)$ as a function of number of drive
cycles $n_0$ for $\omega_D=1$ (top left panel), $\omega_D=0.25$ (top
right panel) and $\omega_D=0.15$ (bottom left panel) indicating the
system size independence of the data for $n_0 \le 75$. Bottom right
panel: Plot of $N_{\rm av}^{(50}\equiv N_{\rm av}(50 T)$ as a
function of $\omega_D$ showing the crossover from delocalized to
localized region. All energies and frequencies are measured in units
of ${\mathcal J}_0$, $\hbar$ is set to unity, and the chain length
is $L=16$. See text for details.} \label{fig8}
\end{figure}

\section{Discussion}
\label{diss}

In this work, we have analyzed a weakly interacting finite chain
subjected to a continuous drive. We have charted out a Floquet
perturbation theory for systematic computation of its Floquet
Hamiltonian. We find that the results obtained from such a
perturbative procedure provides accurate description of the system
dynamics for $\hbar \omega_D \simeq V_0 \ll {\mathcal J}_0$. We note
that in contrast, the Floquet Hamiltonian obtained from Magnus
expansion yields quantitatively accurate results only for $\hbar
\omega_D > {\mathcal J}_0$.

We note that for continually driven systems, the computation of $U$
via exact numerics is difficult since it requires numerical
implementation of time ordering. This usually requires
trotterization of $U$ at infinitesimal time slice $\delta=T/N$. The
computational time for this numerical procedure scales as $2 N
D^{a}$ for $N \gg 1$, where $D= 2^L$ is the Hilbert space dimension
for a chain of length $L$ while the exponent $2 \le a \le 3$ depends
on the choice of algorithm for multiplication of unitary matrices.
In addition this procedure requires an additional $\sim D^{b}$ time
where $b \sim 3 $ for diagonalization of the final unitary matrix.
In contrast finding eigenvalues and eigenvectors of $U$ via FPT
involves two steps. The first involves construction of the Floquet
Hamiltonian $H_F$ using Eqs.\ \ref{fham1} and \ref{fham2}; the
computational time here scales as $n_{\rm max} D^a$ where $n_{\rm
max}$ is the maximum index of Bessel functions that one keeps in the
sum while evaluating the sum in Eq.\ \ref{fham2}. We find that
$n_{\rm max} \sim 5$ is usually enough to obtain accurate results
using second order FPT. The second constitutes diagonalization of
the matrix obtained for $H_F$; in this case, it involves
diagonalization of a hermitian matrix and hence requires ${\rm
O}(D^2)$ computation time. Thus FPT is faster by at least a factor
of $2N/n_{\rm max} \gg 1$ for large $D$ and $N$. This allows us to
numerically obtained spectrum of $H_F$ for $L \le 16$; in contrast,
analogous computation for exact $H_F$ can not be done with same
computational resources   for $L > 12$. We note that
whereas computation of local correlation functions can be carried
out numerically for larger systems, quantities such as the Shannon
entropy $S$ which requires knowledge of eigenvectors of $U$ can not
be easily accessed in these systems without using FPT. Moreover, our
method could allow one, in principle, to access $L\sim 22$ using
cluster computation coupled with techniques to calculate the matrix
elements of $H_F$ on the fly; we leave this as a possible subject of
future work.

Our results indicate that the approach of such driven system to
steady state is accurately captured by FPT. To this end, we compute
$Q$ for an initial thermal mixed state and a product state; for both
of these we find that for finite chain there is a distinct
crossover. For high drive frequency, the system barely evolves and
$Q=-1$ while at low enough frequencies it goes to the ETH predicted
infinite temperature steady state leading to $Q=-1$. In between, for
a distinct range of frequencies, the steady state of a finite chain
assumes either subthermal or superthermal values for $\langle H_{\rm
av}\rangle$ depending on the initial state. A similar feature is
also seen in behavior of $S$. Moreover, the protocol that we use for
driven fermion chain ensures that the non-interacting fermions
exhibit exact dynamical localization at $t_0=n_0T$. Our work
demonstrates that for driven finite interacting chains, the steady
states can be either localized or delocalized; we find a frequency
induced crossover between them around $\hbar \omega_D \simeq
{\mathcal J}_0/2 \gg V_0$. We relate this behavior to the change in
Floquet eigenstates of the driven system.

The implication of our results for thermodynamic large chains can be
understood as follows. For such driven chains, the steady state is
expected to be the ETH predicted infinite temperature state.
However, we note that the system would take a much larger time to
reach such a steady state at high frequencies (where dynamical
localization ensures that such times would be $\sim \exp[a
\omega_D]$ where $a$ is a typical ${\rm O}(1)$ number). In contrast,
for low drive frequencies, the system reaches the steady states
fast, usually within a few drive cycles. Moreover, as shown in Fig.\
\ref{fig8}, numerically using ED, we find that for all system sizes
$L \le 15$ and for representative frequencies shown, the value
$N_{\rm av}(n_0T)$ starting from $|\psi_p\rangle$ almost coincides
for $n_0 \le 75$. This allows us to believe that the behavior of
$N_{\rm av}^{50} \equiv N_{\rm av}(50 T)$ found in these
finite-sized chains would also be seen in thermodynamically large
chains. This behavior is shown in the bottom right panel of Fig.\
\ref{fig8}; we find that $N_{\rm av}^{50}$ closely mimics the steady
state behavior of $N_{\rm av}$ for finite chain. This phenomenon is
a consequence of the fact that the driven chain takes longer to
reach its steady state at higher drive frequencies.

The experimental realization of our work can be done using a
Fermi-Hubbard chain in the weak interaction limit \cite{exp1}. Here
we suggest that the kinetic energy term be made time dependent. This
can be done by subjecting the system to a laser whose intensity
varies with time. Our prediction for finite chain is that the
heating rate of the system as a function of the drive frequency
would exhibit a crossover as seen for $Q$. Moreover one can prepare
such a chain in an initial state $|\psi_p\rangle$ and study the
density profile as a function of the drive frequency. We expect such
a profile to remain localized for high drive frequency and
delocalize for low drive frequencies as shown in the bottom right
panel of Fig.\ \ref{fig7}.

In conclusion, we have studied a continuously driven finite
interacting fermion chain in the weak interaction limit and derived
a Floquet Hamiltonian for the system using FPT. Our analysis
indicate that the FPT works well for $\hbar \omega_D \ge V_0$
allowing access to the dynamics of the system over a wider range of
drive frequencies compared to Magnus expansion. We have studied
steady states of such finite driven chains and their crossover
between dynamically localized to delocalized behavior and discussed
experiments which can test our theory.

\begin{acknowledgments}
R.G. acknowledges CSIR SPM fellowship for support and the authors thank A. Sen for discussion.
\end{acknowledgments}

\end{document}